\documentclass[11pt]{article}

\usepackage[a4paper,margin=1in]{geometry}
\usepackage{amsmath,amssymb,amsthm}
\usepackage{mathtools}
\usepackage{bm}
\usepackage{graphicx}
\usepackage{xcolor}
\usepackage{booktabs}
\usepackage[colorlinks=true, allcolors=blue]{hyperref}
\usepackage[capitalise,nameinlink]{cleveref}
\usepackage{tikz}
\usepackage{algorithm}
\usepackage{algpseudocode}
\usetikzlibrary{arrows.meta,positioning,calc,decorations.pathmorphing,backgrounds,fit,shapes.geometric}
\graphicspath{{figures/}}

\usepackage{enumitem}
\usepackage[numbers]{natbib}
\newcommand{\R}{\mathbb{R}}
\newcommand{\E}{\mathbb{E}}
\newcommand{\diff}{\mathrm{d}}
\newcommand{\eqdef}{\coloneqq}
\newcommand{\Te}{T_{\!e}}
\newcommand{\nA}{n_{\!A}}
\newcommand{\Ne}{N}
\newcommand{\Wsc}{W}            
\newcommand{\Ntilde}{\widetilde{N}}
\newcommand{\Lat}{\bm{L}}
\newcommand{\Bsp}{\bm{B}}
\newcommand{\wsp}{\bm{w}}
\newcommand{\Rate}{R_{L}}
\newcommand{\Fth}{F_\theta}
\newcommand{\fE}{f_E}
\newcommand{\fD}{f_D}
\newcommand{\Leq}{\Lat_{\mathrm{eq}}}
\newcommand{\Trans}[1]{\mathcal{T}(#1)}

\usepackage[capitalise,nameinlink]{cleveref}

\title{\bfseries A conservative latent neural ODE for stiff collisional--radiative plasma kinetics}

\author{
  Xuping Xie\textsuperscript{1}\thanks{Corresponding author: \texttt{xxie@odu.edu}}
  \quad
  Xian-Zhu Tang\textsuperscript{2}
  \quad
  Qi Tang\textsuperscript{3}
  \\[0.6em]
  \begin{minipage}{0.95\textwidth}
    \centering\footnotesize\itshape
    \textsuperscript{1}Department of Mathematics \& Statistics, Old Dominion University, Norfolk, VA, USA\\[0.15em]
    \textsuperscript{2}Theoretical Division, Los Alamos National Laboratory, Los Alamos, NM, USA\\[0.15em]
    \textsuperscript{3}School of Computational Science and Engineering, Georgia Institute of Technology, Atlanta, GA, USA
  \end{minipage}
}
\date{}

\begin{document}
\maketitle

\vspace{-2em}
\begin{abstract}
\noindent
Collisional--radiative (CR) models govern the non-equilibrium population
kinetics of partially ionized plasmas and are a dominant cost in integrated
edge- and divertor-plasma simulations, where a stiff system of dozens to
thousands of coupled rate equations must be solved at every spatial cell and
time step. We present a two-stage machine learning (ML) surrogate that replaces the
direct CR solve with a compact, continuous-time, deployable dynamical model.
Stage~1 is a grey-box autoencoder that compresses a high-dimensional ion
population vector into a low-dimensional latent space in which four  coordinates, one per charge state of the lithium system studied here, are
\emph{fixed} to the physical charge-state distribution and the rest are learned.
Stage~2 is a latent neural ODE that learns the
continuous-time evolution of the latent variables. We (i) integrate in a density-folded, logarithmic clock on
which the latent trajectory is smooth enough for an explicit integrator,
(ii) sample time adaptively by latent arc length so the sharp ionization
transition is resolved, (iii) add a linear damping toward an estimate of the
CR steady state as a design choice, and (iv) enforce mass
conservation directly. The surrogate is accurate, with a charge-state error that is unchanged under refinement of its output grid, over
full-trajectory rollouts within the trained horizon, and exactly conservative: on $1480$ held-out
trajectories at the trained conditions it reproduces the charge-state evolution with a median absolute
error of $4.2\times10^{-3}$ on fractions in $[0,1]$, conserves mass to
$10^{-7}$, {and evaluates a full sixteen-decade trajectory in ${\sim}76\,\mu$s per trajectory in batched inference on a single GPU.} A model retrained with whole $(\nA,\Te)$ grid points withheld is equally accurate at those unseen conditions (median $3.8\times10^{-3}$).
\end{abstract}

\section{Introduction}
\label{sec:intro}
In a partially ionized plasma that is neither in local thermodynamic equilibrium
nor in coronal equilibrium, the populations of atomic and ionic energy levels are
governed by the competition between collisional processes (electron-impact
excitation, de-excitation, and ionization, and collisional (three-body)
recombination), radiative
processes (spontaneous emission, radiative recombination), and dielectronic
recombination, a resonant capture stabilized by photon emission
\citep{capitelli2012fundamental,celiberto2017elementary}.
The collisional--radiative (CR) model \citep{Bates1962} encodes this competition as a
system of coupled ordinary differential equations for the level population
vector $\Ne(t)\in\R^{n}$,
\begin{equation}
\frac{\diff \Ne}{\diff t} = \bm{M}(\Ne;\nA,\Te)\,\Ne,
\label{eq:cr}
\end{equation}
where $\bm{M}$ collects the rate coefficients at total species density
$\nA$ and temperature $\Te$. {The electron density $n_e$ entering the
  coefficients is set by quasineutrality from the instantaneous
  charge-state densities, so $\bm{M}$ depends on the state $\Ne$ as
  well as on the conditions $(\nA,\Te)$, which fix the
  background. Since $n_e$ is linear in $\Ne$, the terms driven by a single
  electron (electron-impact processes, radiative and dielectronic
  recombination) are quadratic in $\Ne$, and collisional recombination,
  which scales as $n_e^{2}$, is cubic; at fixed conditions
  \eqref{eq:cr} is therefore autonomous but polynomially nonlinear in $\Ne$;
  every trajectory nonetheless relaxes to a steady state, referred to
  below as the CR steady state (\Cref{sec:analysis}).} 
The dimension $n$ ranges from tens to many
thousands of levels depending on the resolution (superconfiguration,
configuration-averaged, or fine-structure) \citep{Summers2006}; the lithium
model treated in this paper resolves $n=94$ levels across four charge states. CR models
generalize the equilibrium ionization balance long tabulated for
astrophysical plasmas \citep{arnaud1992iron} to finite density and to
time-dependent conditions, and they supply the line emission and
radiative power loss that close the equations for impurity-transport and plasma energy
evolution in magnetic-confinement fusion; they are likewise central to
spectroscopic diagnostics and to opacity and cooling-curve
calculations \citep{Chung2005FLYCHK}.

Equation~\eqref{eq:cr} is severely \emph{stiff}: the real parts of the
eigenvalues of {the instantaneous} $\bm{M}$ differ by many orders of
magnitude, because collisional and radiative rates differ greatly in
size and scale differently with $\nA$ and $\Te$. Stiffness of this
kind is characteristic of CR and also of chemical kinetics in general,
and is the primary obstacle to building learned surrogates for such
systems \citep{Ji2021,Kim2021Stiff,Goswami2024,nockolds2026lilan}.
Resolving the fast modes requires implicit integration or small time
steps, and {because the system is nonlinear each implicit step is a
  Newton iteration, every pass of which solves a linear system in the
  level populations, an $O(n^{3})$ factorization repeated several
  times per step for $n$ up to thousands of levels.} When a CR solve
is embedded in an integrated transport code, such as SOLPS-ITER
\citep{Wiesen2015}, or fluid/gyrokinetic edge codes, it must be
evaluated at every spatial cell and time step across a wide range of
local $(\nA,\Te)$ conditions, so the aggregate CR cost can dominate
the simulation. This motivates fast surrogates that reproduce the CR
dynamics at a small fraction of the cost while remaining accurate and
physically consistent enough to be trusted inside a multi-physics
simulation. We will demonstrate our surrogate modeling approach with
the case study of lithium, a low-$Z$ element of growing interest as a
liquid plasma-facing material, whose CR kinetics govern its ionization
and radiation in the edge plasma \citep{deCastro2021}.

Deep learning has recently advanced core fusion tasks from disruption
prediction to plasma control \citep{KatesHarbeck2019,Degrave2022}, and
physics-informed learning provides a general framework for embedding physical
structure in such models \citep{Karniadakis2021}. For CR kinetics specifically,
machine-learning (ML) surrogates are attractive because the CR solution manifold
is low-dimensional in practice{: in this work, the $94$ levels of a
	single lithium species compress to a ten-dimensional latent space} (\Cref{sec:analysis}). Neural-network surrogates
for CR models have been developed with adaptive sampling of the parameter space
\citep{garland2020progress,garland2022efficient}. A prior two-stage approach
\citep{Xie2024} exploited this structure directly: a physics-assisted
autoencoder compresses the population vector into a compact latent variable, which a
discrete-time flow-map network (FMNet) then advances by a fixed step. That work
established the viability of the latent-surrogate idea but left three
limitations that matter for deployment: (i) the flow map is tied to the training
time step, so the surrogate is not readily usable at the variable steps an
integrated code takes;
(ii) errors accumulate over long rollouts; and (iii) mass conservation
is imposed only softly, through a loss penalty, so the predicted populations
may not sum to unity. The surrogate developed here addresses all three: it is
continuous in time, so the output times along a trajectory are a free choice
rather than a property fixed at training (advancing an arbitrary intermediate
state is a separate matter, discussed under Limitations), its integrated field
includes a linear damping toward an estimate of the CR steady state,
and it conserves mass exactly by construction. {A full trajectory takes ${\sim}76\,\mu\mathrm{s}$ on a single
GPU when trajectories are evaluated in batches (\Cref{sec:setup}), a throughput compatible with in-loop use.}

\paragraph{Contributions.}
This paper contributes a continuous-time surrogate for stiff CR
kinetics, built on two ideas, and validates it on held-out trajectories across
the parameter grid, at conditions withheld from training, and against a flow-map baseline from \citet{Xie2024}, retrained
on the identical data, split, and latent representation (\Cref{sec:results}).
The first is a \emph{physics-based time reparametrization under which explicit
integration of the latent dynamics becomes practical}: the latent dynamics are integrated not in physical time but in a
density-folded logarithmic clock, $\tau\propto\log_{10}(t\,\nA)$, which maps the
sixteen decades of physical time onto a unit interval on which the latent
trajectory is smooth except for one localized transition, and absorbs the
density-dependent shift of that transition. The reparametrization rescales the
physical Jacobian by a common factor and therefore does not by itself remove
stiffness; it is the combination of the latent reduction, the clock, and
arc-length-adaptive sampling that permits accurate explicit integration on the
tested trajectories, in about a hundred Runge--Kutta steps (\Cref{sec:setup}). The second is an \emph{exactly conservative dynamical model}: the latent dynamics are learned as a conditioned
neural ODE whose integrated field includes a linear damping toward an estimate of the
CR steady state {(regressed on $(\nA,\Te)$ from the simulated terminal states;
	\Cref{sec:vf})}, a design choice that gave no measurable accuracy gain
in an ablation (\Cref{sec:res-deploy}),
and whose decoder enforces
mass conservation exactly through a simplex projection, the positive
normalization of \Cref{eq:simplex}. 

The remainder of the paper is organized as follows.
\Cref{sec:related} places the surrogate in the context of learned reduced models
for stiff kinetics, neural differential equations and their stabilization, and
the exact enforcement of physical constraints. \Cref{sec:approach} states the
problem, develops the grey-box autoencoder and the latent neural ODE,
and then characterizes the structural properties of the data that each design
choice exploits.
\Cref{sec:results} evaluates the surrogate on held-out trajectories
and against the retrained flow-map baseline, and \Cref{sec:conclusions}
concludes.
\section{Related Work}
\label{sec:related}

Our surrogate sits at the intersection of three lines of work: learned reduced
models for stiff kinetics, neural differential equations, and
structure-preserving ML with strongly enforced physical constraints; we close by
contrasting the latent-dynamics route with operator learning. Surrogates for
stiff chemical kinetics have been built as neural ODEs \citep{Owoyele2022} and
as physics-informed neural networks \citep{Ji2021}, while
reduced models of stiff dynamics more generally combine autoencoder-based
coordinate reduction \citep{Champion2019} with discrete-time flow maps or direct
state-to-state regression. {Relative to the CR surrogate of~\citet{Xie2024}
described in \Cref{sec:intro}, we keep the grey-box autoencoder and replace only
the flow map.} Recent high-fidelity CR modeling for
fusion-relevant species \citep{sharma2026hybrid,sharma2026collisional} would
supply the trajectory ensembles needed to extend surrogates of this type beyond
lithium.

The dynamical core builds on neural ODEs, which parametrize the right-hand side
of an ODE with a neural network and optimize it by differentiating the solve,
either directly or through a time-reverse adjoint system~\citep{Chen2018NODE},
and on the latent-ODE variant that learns the dynamics in a learned latent space
and is well suited to irregularly sampled series \citep{Rubanova2019}. Training
such models on stiff systems is challenging, because both the forward and
adjoint solves inherit the stiffness \citep{Kim2021Stiff}; the principal
remedies are equation scaling and time reparametrization, which transform the
integration variable so the fastest mode becomes $\mathcal{O}(1)$
\citep{Caldana2025}, and architectural or gradient regularization that improves
conditioning \citep{Fronk2025}. Our density-folded logarithmic clock is a
\emph{physics-defined} reparametrization in this family; we show that, combined
with the latent reduction and arc-length sampling, it permits an explicit
Runge--Kutta solver to integrate the lithium system accurately in about a hundred
steps per trajectory (\Cref{sec:setup}). A separate
difficulty is that long autonomous rollouts of learned dynamics tend to drift or
diverge. A now-standard fix adds a linear damping
term whose fixed point anchors the long-time solution. \citet{Linot2023} learn
such a linear operator jointly with the nonlinear vector field and show it keeps
trajectories on the attractor for chaotic and shock-laden systems. We adopt this
damping term but specialize it to our setting: {rather than letting the
spectrum of a learned operator define the attractor, we damp toward an estimate of the CR steady state, regressed on
$(\nA,\Te)$ from the terminal states of the simulated trajectories.} In our
trained model the damping does not make the terminal state attracting, and a
model trained without it is as accurate (\Cref{sec:res-deploy,sec:res-field}),
so here it is a design choice rather than a demonstrated stabilizer. Closely related in spirit,
physics-informed latent dynamics identification with stability and steady-state
constraints has recently been applied to time-dependent NLTE atomic
kinetics~\citep{Nam_2026}, sharing our premise
that atomic-kinetics surrogates should model the time evolution and be anchored
to the correct equilibrium.

We also draw on work that enforces conservation laws \emph{exactly} rather
than through soft penalties. In combustion and reacting-flow networks, a
projection or constraint layer can guarantee element and mass conservation
\citep{Readshaw2023,ANNhard2025}; we adopt an analogous simplex projection{, applied to the level-resolved
output, whose partial sums then give charge-state fractions that satisfy the
constraint identically (\Cref{sec:aearch})}. A numerically stable logarithmic conservation
penalty is also kept in both training stages, so that the learned representation
and dynamics stay close to the constraint and the projection has little to
correct.

A different formulation bypasses latent dynamics altogether and learns the
solution operator mapping conditions and time directly to the state. Such
operator-learning surrogates have been demonstrated on stiff chemical kinetics
\citep{Goswami2024}, and a linear latent
network that integrates its latent dynamics analytically has likewise been
demonstrated as a real-time solution operator for stiff kinetics, including a
lithium CR model \citep{nockolds2026lilan}. We pursue the latent-dynamics route
because it yields an interpretable, conservable, continuous-time
integrator. We do not attempt a head-to-head comparison against these
operator-learning surrogates: a meaningful one would require running them on
identical trajectory data, and no common benchmark yet exists for CR kinetics.
Our quantitative baseline is instead the discrete flow-map surrogate
of~\citet{Xie2024}, retrained here on the identical data, split, and latent
representation (\Cref{sec:results}), which isolates the effect of the
continuous-time formulation.

\section{Proposed Approach}
\label{sec:approach}

This work focuses on the lithium CR system of \eqref{eq:cr}, for which we follow
the notation of~\citet{Xie2024}. The state $\Ne(t)\in\R^{94}$ is the population
density of lithium, in $\mathrm{m}^{-3}$, over the ground and excited states of
the four charge states $n_0$ (neutral) through $n_3$ (fully stripped). The
background is a Maxwellian electron distribution of temperature $\Te$ (in eV)
together with the total lithium density $\nA=\sum_i\Ne_i$; {quasineutrality sets
the electron density from the instantaneous charge-state densities, so $\bm{M}$
depends on the state as well as on $(\nA,\Te)$, which fix the background.} Because $\Ne$ spans many decades across conditions we
normalize the density populations $\Ntilde=\Ne/\nA$, whose group sums
over the levels of each charge state give the charge-state fractions
$n_z\in\R^{4}$, with $\sum_i n_{z,i}=1$ by construction. The remaining output is
the radiative power loss $\Rate(t)$, the rate at which the electrons lose energy
to radiation. Given $(\nA,\Te)$ and an initial
state, the surrogate must reproduce $\Ne(t)$ and, from it, the $n_z(t)$ and
$\Rate(t)$ that downstream transport and radiation models consume.

We describe the two stages in turn, the grey-box autoencoder that learns the
latent representation (\Cref{sec:ae}) and the latent neural ODE that
advances it (\Cref{sec:node}), and then characterize in \Cref{sec:analysis}
the properties of the data that each design choice exploits.

\subsection{Grey-box autoencoder}
\label{sec:ae}

\paragraph{Standardization.}\label{sec:scaling}
Every quantity is positive and spans many decades, so each is mapped to a
standardized coordinate by a per-channel base-ten logarithm followed by a
$z$-score. With $\bm\mu,\bm\sigma$ the per-channel log-mean and
log-standard-deviation:
\begin{align}
\text{populations:}\quad &
\Wsc=\Trans{\Ne}\eqdef\big(\log_{10}\max(\Ntilde,\varepsilon_N)-\bm\mu_N\big)\oslash\bm\sigma_N,
\label{eq:scaleN}\\
\text{charge states:}\quad &
\wsp_z=\big(\log_{10}\max(n_z,\varepsilon_N)-\bm\mu_z\big)\oslash\bm\sigma_z,
\label{eq:scalenz}\\
\text{radiative loss:}\quad &
r=\big(\log_{10}\max(\Rate,\varepsilon_R)-\mu_R\big)/\sigma_R,
\label{eq:scaleR}\\
\text{conditions:}\quad &
c=(\log_{10}\nA,\ \Te)\ \text{each $z$-scored},
\label{eq:scaleP}
\end{align}
where $\oslash$ is elementwise division and $\varepsilon_N,\varepsilon_R${ ($=10^{-30}$, far below any physical population, so the floors are inert except at exact zeros)} floor the
logarithm. The logarithm absorbs the dynamic range and the $z$-score leaves every
channel at $\mathcal{O}(1)$, so no single channel dominates the loss. Each
transform inverts in closed form, e.g.\
$\Ntilde=10^{\,\Wsc\odot\bm\sigma_N+\bm\mu_N}$. The constants are accumulated in one
streaming pass and reused unchanged at training, evaluation, and deployment.

\paragraph{Architecture.}\label{sec:aearch}
The autoencoder compresses $\Wsc\in\R^{94}$ into a ten-dimensional latent space,
four of whose coordinates are fixed by physics rather than learned:
\begin{equation}
\Lat=[\,\wsp_z,\;\Bsp\,],\qquad
\wsp_z\in\R^{4}\ (\text{white, fixed}),\quad \Bsp=\fE(\Wsc)\ (\text{black, learned}).
\label{eq:greylatent}
\end{equation}
The white block is the scaled charge-state vector $\wsp_z$ of \eqref{eq:scalenz},
supplied directly; the black block is the encoder output, with $\fE$ an MLP. The
decoder maps the full latent variable back to populations and radiative loss,
$[\widehat{\Wsc},\,\widehat{r}\,]=\fD(\Lat)$. This is what makes the representation a
grey box. The encoder is left with only the six {learned coordinates} to
find, and the four physical coordinates remain available downstream, where the
neural ODE supervises them directly.
{The decoder produces the level populations in $\log_{10}$ space, so the raw
physical reconstruction
$\Ntilde^{\mathrm{raw}}=\mathcal{T}^{-1}(\widehat{\Wsc})
=10^{\,\widehat{\Wsc}\odot\bm\sigma_N+\bm\mu_N}$ is strictly positive by
construction; the simplex projection is then the normalization
\begin{equation}
\Ntilde_i=\Ntilde^{\mathrm{raw}}_i\Big/\textstyle\sum_{j}\Ntilde^{\mathrm{raw}}_j,
\label{eq:simplex}
\end{equation}
which places the level-resolved output on
$\{\Ntilde_i>0,\ \sum_i\Ntilde_i=1\}$ to floating-point rounding. Unlike a
Euclidean projection it never truncates a small population to exactly zero, and
it preserves the ratios of populations spanning many decades; the charge-state
fractions are partial sums of $\Ntilde$ and therefore inherit
$\sum_i n_{z,i}=1$ identically.}

\paragraph{Training objective.}
The autoencoder minimizes, over all trajectory time points,
\begin{equation}
\mathcal{L}^{\mathrm{AE}}
= w_W\,\mathcal{L}_W + w_R\,\mathcal{L}_R  + w_m\,{\mathcal{L}_{\Sigma N}},
\label{eq:aeloss}
\end{equation}
where
\begin{align}
\mathcal{L}_W = \lVert \widehat{\Wsc}-\Wsc\rVert_2^2, \qquad
\mathcal{L}_R = (\widehat{r}-r)^2, \qquad
{\mathcal{L}_{\Sigma N}} = \big(\log\textstyle\sum_i\Ntilde_i^{\mathrm{raw}}\big)^2,
\end{align}
with $\Ntilde^{\mathrm{raw}}=\mathcal{T}^{-1}(\widehat{\Wsc})$ the reconstruction
returned to physical units. Both $(\log\sum\Ntilde)^2$ and the raw
$(\sum\Ntilde-1)^2$ vanish at $\sum\Ntilde=1$; we use the logarithmic form because
it does not amplify the error of the exponentiated inverse transform early in
training, when $\sum\Ntilde$ can still be far from one.

\begin{figure}[htp]
\centering
\includegraphics[width=\linewidth]{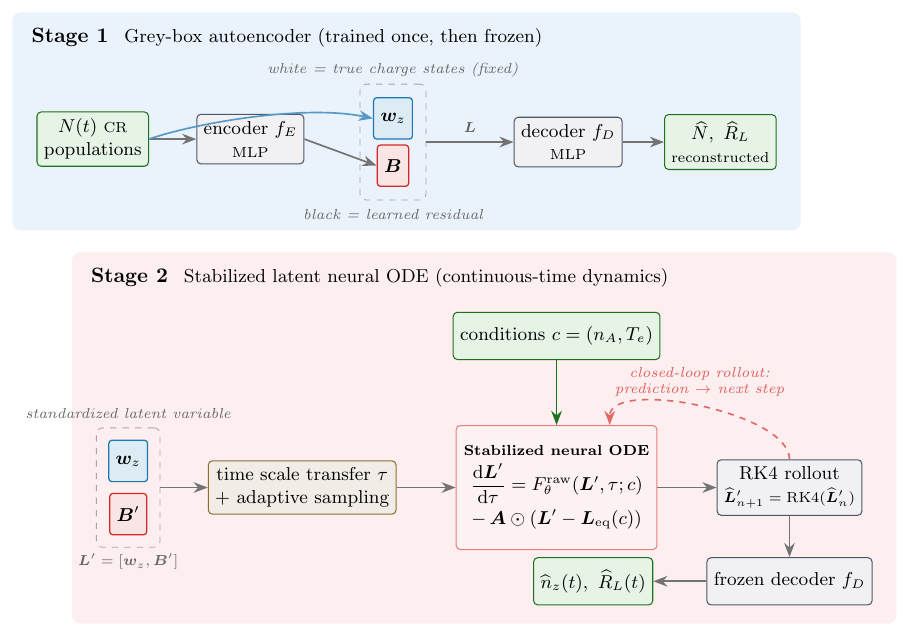}
\caption{The two-stage surrogate. \textbf{Stage 1}: a grey-box autoencoder
compresses the scaled $94$-level population vector into a ten-dimensional latent space
whose four \emph{white} coordinates are the true charge-state distribution, fed
in directly (blue), and whose six \emph{black} coordinates are learned (red);
it is trained once and frozen. \textbf{Stage 2}: the black block is
standardized, and the full latent variable $\bm{L}'=[\wsp_z,\Bsp']$ is advanced
in the density-folded clock $\tau$ by a
latent neural ODE (a learned field $F_\theta^{\mathrm{raw}}$ plus linear damping
toward the CR steady state $\bm{L}_{\mathrm{eq}}(c)$), integrated by
explicit RK4. The frozen decoder returns charge states and radiative loss, with
a simplex projection (not shown) enforcing mass conservation.}
\label{fig:architecture}
\end{figure}

\subsection[Latent neural ODE]{Latent neural ODE}
\label{sec:node}
Stage~2 learns the continuous-time evolution of the latent variable. All
trajectories are encoded once with the frozen autoencoder; the neural ODE trains
on those latent variables
and never revisits the raw populations. The white block $\wsp_z$ is already
standardized, but the black block $\Bsp$ is the encoder's raw output, whose
per-coordinate scale is uncontrolled. We standardize it with the per-coordinate mean and
standard deviation of the encoder outputs over the \emph{training} trajectories,
\begin{equation}
	\Bsp'=(\Bsp-\bm\mu_{\Bsp})\oslash\bm\sigma_{\Bsp},\qquad \Lat'=[\,\wsp_z,\;\Bsp'\,]\in\R^{10},
	\label{eq:blackstd}
\end{equation}
so all ten coordinates are $\mathcal{O}(1)$ and neither block dominates the latent
loss. The neural ODE works in $\Lat'$ throughout; $\Bsp'$ is mapped back to $\Bsp$
only when a prediction is passed to the decoder.

\paragraph{Why a learned latent field.}
The latent field is learned from trajectories, not derived from \eqref{eq:cr}. No
derivative labels are supplied: the network predicts a field, the field is
integrated, and the parameters are adjusted to reduce the error of the
integrated trajectory against the encoded CR states at the saved times. The
alternative is to transform the physical equations through the encoder: if
$\Lat'=E(\Ne)$ denotes the complete encoding, the chain rule gives
\begin{equation}
\frac{\diff\Lat'}{\diff\tau}=\frac{\diff t}{\diff\tau}\,DE(\Ne)\,\bm{M}(\Ne;c)\,\Ne,
\label{eq:chainrule}
\end{equation}
with $DE$ the Jacobian of the encoding. Evaluating this expression requires the full
CR right-hand side at a decoded state at every step, which is precisely the cost
the surrogate is meant to remove, and it is exact only if physically distinct
states with the same encoding have the same encoded derivative, a closure that
no fixed compression guarantees. We therefore learn the field directly and
treat it as a predictive model over the sampled distribution of trajectories:
accurate held-out rollouts support that use, but do not identify the field
uniquely away from the data or establish reliable evolution from arbitrary
latent states (\Cref{sec:conclusions}).

\paragraph{Time reparametrization.}
\label{sec:clock}
The model is a parametrized family of dynamics conditioned on $c$
in \eqref{eq:scaleP}. It integrates in a \emph{density-folded logarithmic clock}
$\tau$ rather than in physical time, for two reasons. The kinetics are stiff in
$t$, spanning sixteen decades with the charge-state rate {per second varying by
roughly eight orders of magnitude} along one trajectory, but smooth per decade of $\log_{10}t$: a
logarithmic clock stretches those decades into a unit interval on which the
latent trajectory is smooth apart from one localized transition, so that a
fixed budget of about a hundred steps can be placed where the state moves. The transition also moves with density,
$t_{\mathrm{trans}}\propto\nA^{-1}$, so a plain $\log_{10}t$ clock would place it
somewhere different for every density and the vector field would have to learn
that shift. Multiplying time by $\nA$ before taking the logarithm holds
$t\,\nA$ at the transition roughly constant, which brings the transitions much
closer together on the clock. Normalizing to the unit interval,
\begin{equation}
	\tau=\frac{\log_{10}(t\,\nA)-a}{b-a}\in[0,1],
	\qquad
	t=\frac{10^{\,\tau(b-a)+a}}{\nA},
	\label{eq:tau}
\end{equation}
where $a$ and $b$ are the dataset extrema of $\log_{10}(t\,\nA)$; the second
expression is the exact inverse, used to map predictions back onto physical time.
Writing the physical system at fixed conditions as $\dot{\Ne}=f(\Ne;c)$ with
$f=\bm{M}\Ne$, the change of variable gives
\begin{equation}
	\frac{\diff\Ne}{\diff\tau}=\alpha(\tau)\,f(\Ne;c),\qquad
	\alpha(\tau)\eqdef\frac{\diff t}{\diff\tau}=(b-a)\ln(10)\,t(\tau),
	\label{eq:alpha}
\end{equation}
so the transformed system carries an explicit, known clock dependence through
$\alpha(\tau)$ although the atomic coefficients have none. Two consequences
follow. First, the clock input to the learned field (\Cref{sec:vf}) exists to
represent this induced dependence; a more constrained architecture,
$\diff\Lat'/\diff\tau=\alpha(\tau)\,g_\theta(\Lat';c)$, would impose it exactly
and is a natural variant to investigate. Second, at fixed $\tau$ the Jacobian
of the transformed system is $\alpha(\tau)$ times that of the physical system,
so the reparametrization rescales all eigenvalues by a common factor and leaves
their ratios unchanged: it redistributes resolution along the trajectory but
does not remove stiffness in the spectral sense (\Cref{sec:integrator}).
Where no such physical clock is available, a learned reparametrization
\citep{Caldana2025} is an alternative.

\begin{figure}[t]
	\centering
	\includegraphics[width=\linewidth]{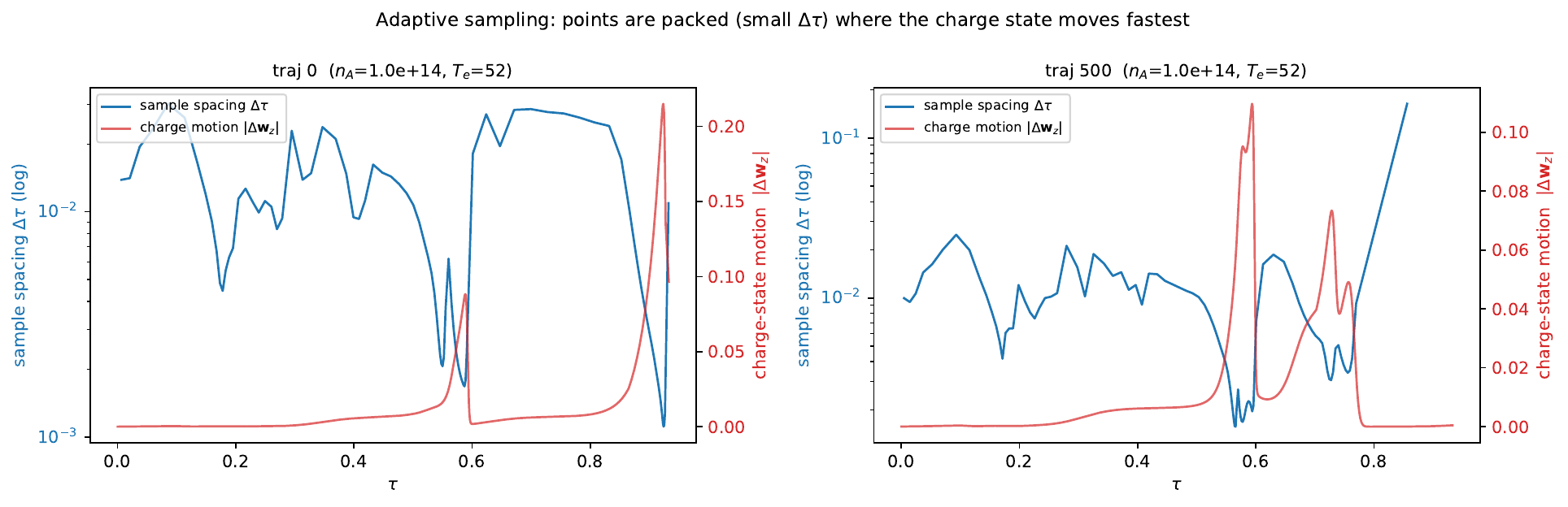}
	\caption{Adaptive arc-length sampling on two representative trajectories. The
		per-step spacing $\Delta\tau$ (blue, logarithmic left axis) collapses where the
		charge-state motion $\lVert\Delta\wsp_z\rVert$ (red, right axis) peaks, so the
		sampler refines the ionization transition and coarsens the plateaus. A
		uniform-in-$\tau$ grid would instead spread points evenly and leave the
		transition under-resolved.}
	\label{fig:adaptive}
\end{figure}

\paragraph{Adaptive arc-length sampling.}
\label{sec:sampling}
Even on the folded clock the ionization transition occupies a median $17\%$ of
the $\tau$ range, and only $0.3\%$ for the sharpest trajectories (\Cref{sec:analysis}). Uniform sampling would therefore
place few points, sometimes none, in the transition region the surrogate must resolve. We
propose to place each trajectory's $n_\tau$ points by the cumulative arc length of
its latent path,
\begin{equation}
	s(\tau)=\int_{\tau_0}^{\tau}\Big\lVert \frac{\diff\Lat'}{\diff\tau'}\Big\rVert\,\diff\tau',
	\qquad
	\text{density}(\tau)\ \propto\ \big\lVert \diff\Lat'/\diff\tau\big\rVert+\epsilon_{\mathrm{arc}}\bar\rho,
	\label{eq:arclen}
\end{equation}
spreading points uniformly in $s$ and mapping back to $\tau$, with a small
uniform floor $\epsilon_{\mathrm{arc}}$ (mean density $\bar\rho$) so flat regions
keep some samples. The integrand is the local rate of change of the state, so the
grid is dense where the state moves quickly and coarse across the plateaus.
The grid points are obtained by inverting the cumulative arc length of the
saved trajectory and therefore fall between saved times; the reference latent on
the grid is the linear interpolation in $\tau$ of the standardized reference
latent between the two neighbouring saved samples, with the first and last grid
points pinned to the first and last saved times. The floor is
$\epsilon_{\mathrm{arc}}=0.05$. The same
construction is used at inference: every rollout reported in
\Cref{sec:results}, including the timing and the grid-refinement study,
integrates on an arc-length grid built from the reference trajectory's own
latent path. Such a grid supplies the timing of the transition, though no
state values, from the reference solution. \Cref{sec:res-deploy} evaluates grids
that do not require it, including a two-pass grid built from the model's own
first prediction, which needs only the initial state and the conditions and
nearly matches the accuracy of the reference grid at twice the cost.
\Cref{fig:adaptive} shows this on two trajectories: the spacing $\Delta\tau$
collapses where the charge-state motion $\lVert\Delta\wsp_z\rVert$ peaks.
\Cref{sec:analysis} quantifies the gain over uniform sampling across the
dataset.

\paragraph{Vector field and linear damping.}
\label{sec:vf}

The learned vector field is an MLP $G_\theta$ acting on the concatenation of the
latent state, a fixed trigonometric embedding of the clock, and the conditions,
\begin{equation}
	\gamma(\tau)\eqdef\big[\sin(2^{k}\pi\tau),\,\cos(2^{k}\pi\tau)\big]_{k=0}^{K-1}\in\R^{2K},
	\quad
	\Fth^{\mathrm{raw}}(\Lat',\tau;c)\eqdef G_\theta\big(\mathrm{concat}[\,\Lat',\,\gamma(\tau),\,c\,]\big),
	\label{eq:fourier}
\end{equation}
with $K{=}4$, so that at each evaluation of the right-hand side eight prescribed
trigonometric values replace the scalar clock input. Together with the ten latent
coordinates and the two conditions they form twenty inputs, and the output is the
ten-dimensional derivative. There is no Fourier transform, no decomposition of
trajectory windows, and no fitted Fourier coefficient: the frequencies are fixed
and only the MLP weights are trained. The construction is the Fourier-feature
input of \citet{Tancik2020}, used because an MLP reading a scalar coordinate
directly fits its low-frequency dependence first and its rapid variation poorly,
the spectral bias documented by \citet{Rahaman2019}; the motivation is easier
learning of the clock dependence, not a change in the dynamics or a stability
guarantee. The clock dependence to be learned is the induced factor
$\alpha(\tau)$ of \Cref{eq:alpha}, which is smooth: the highest frequency,
$8\pi$, has a period of one quarter of the clock range, so the embedding
represents that smooth dependence and not the ionization transition itself,
which for the sharpest trajectories occupies $0.3\%$ of the range
(\Cref{sec:analysis}). The transition is resolved by the dependence of the field
on the latent state and by the arc-length grid of \Cref{sec:sampling}, which
concentrates the integration steps there; four frequency pairs do not by
themselves guarantee resolution of the narrowest transient. In the single-run
ablation of \Cref{sec:res-deploy}, a model given the scalar clock in place of
the embedding reached a validation loss about twelve times higher, and some of
its test rollouts were not finite.

Following \citet{Linot2023}, the integrated field adds a linear damping toward a
conditioned equilibrium anchor:
\begin{equation}
	\frac{\diff\Lat'}{\diff\tau}=\Fth(\Lat',\tau;c)
	=\Fth^{\mathrm{raw}}(\Lat',\tau;c)-\bm{A}\odot\big(\Lat'-\Leq(c)\big),
	\label{eq:Fimpl}
\end{equation}
where $\bm{A}=\mathrm{softplus}(\bm{a})\in\R_{>0}^{10}$ is a learnable positive
diagonal damping. \citet{Linot2023} learn the linear operator and let its spectrum
define the attractor. Here the anchor is instead estimated from data: {at fixed
	conditions the trajectories in the dataset approach a terminal state
	(median terminal drift ${\sim}2\times10^{-8}$; a few low-temperature
	trajectories are still evolving at the final saved time, \Cref{sec:analysis}), and the dominant charge-state
	fractions of the terminal state depend on $(\nA,\Te)$}, with the singly ionized
stage dominant at low $\Te$ and the fully stripped stage at high $\Te$. Trace
populations, such as the post-transition $n_0$, settle to values that depend on
the initial condition (\Cref{fig:charge-nper}) and stay there to the end of the
simulation at $1\,$s. We take $\Leq(c)$ to be a smooth estimate of
this terminal state{, supplied by a small \emph{anchor network}
$\Leq(c)=\mathrm{MLP}_{\mathrm{eq}}(c)$ regressed against the measured terminal
latent variables (see below).} The damping term adds no capacity that
$\Fth^{\mathrm{raw}}$ lacks: its state-independent part
$\bm{A}\odot\Leq(c)$ is a conditioned offset that the MLP could absorb, and the
state-dependent part $-\bm{A}\odot\Lat'$ is a linear contraction. It is retained
as an inductive bias, following \citet{Linot2023}, intended to favour a
contracting terminal state; in the trained model it does not produce one
(\Cref{sec:res-field}). It is a design
choice, not a demonstrated requirement: in the ablation of \Cref{sec:res-deploy},
a model trained without it ($\bm{A}=0$) has the same median accuracy.

\paragraph{Integration.}
\label{sec:integrator}
Each trajectory is integrated on its non-uniform grid $\{\tau_0<\dots<\tau_M\}$
($M{+}1=n_\tau$) by the classical explicit fourth-order Runge--Kutta scheme with
steps $h_n=\tau_{n+1}-\tau_n$. Training rolls out $H<M$ steps at a time;
inference integrates the whole grid. {Parameter gradients are obtained by
backpropagating directly through the RK4 solve rather than by the continuous
adjoint method; with $n_\tau=100$ the unrolled graph fits comfortably in
memory, and this avoids the second, time-reversed solve that would itself
inherit any residual stiffness.}
Two clarifications about what this integration does and does not establish.
First, the clock does not remove stiffness by itself: by \Cref{eq:alpha} it
rescales the physical Jacobian by $\alpha(\tau)$ and leaves eigenvalue ratios
unchanged. What the explicit scheme integrates is the learned latent field,
whose Jacobian is not that of \eqref{eq:cr}; the ten-dimensional latent
representation does not carry the fast physical modes, the clock spreads the
slow evolution over a unit interval, and the adaptive grid resolves the one
sharp feature. The combined latent reduction, time reparametrization, and
sampling strategy permits accurate explicit integration on the tested
trajectories; we do not claim elimination of the original fast modes, and
this paper does not separate the three contributions by ablation (the
ablation of \Cref{sec:res-deploy} changes the training grid only) or examine
the Jacobian of the complete learned field, so no one of them is credited with
the result. Second, the arc-length grid is an
output-sampling choice, not an error or stability control: each pair of
adjacent grid points is one RK4 step, with no internal substepping or error
estimate, and the state-speed criterion of \Cref{eq:arclen} says nothing about
the stability bound $h\lambda\lesssim2.8$ of classical RK4 for a mode of rate
$\lambda$. Stability and accuracy are therefore verified empirically by grid
refinement (\Cref{sec:res-cons}): the charge-state error is unchanged from
$100$ to $400$ points (the latent error on the late plateau is not, see
\Cref{fig:grid-invariance}), and a $50$-point grid is unstable, which locates the stability boundary
of the learned field on this family of grids. \Cref{sec:res-deploy} also
integrates the same field with an error-controlled scheme and measures the RK4
integration error directly. Larger steps than those trained,
and extrapolation beyond the clock domain, are not justified.

\paragraph{Training objective.}
\label{sec:nodeloss}
Training samples random rollout horizons $H$, drawn up to a maximum that grows
from $2$ to $56$ steps over the course of training (\Cref{sec:setup}). For a
horizon $H$ the model integrates
from a true initial latent variable and is supervised against the encoded ground
truth $\Lat'^{\star}$ through the loss,
\begin{equation}
	\mathcal{L}^{\mathrm{NODE}}
	=\lambda_w\,\mathcal{L}_{\text{white}}+\lambda_b\,\mathcal{L}_{\text{black}}
	+\lambda_{\mathrm{eq}}\,\mathcal{L}_{\mathrm{eq}}+\lambda_c\,{\mathcal{L}_{\Sigma n_z}},
	\label{eq:nodeloss}
\end{equation}
with
$\mathcal{L}_{\text{white}}=\E\lVert\widehat{\wsp}_z-\wsp_z^{\star}\rVert_2^2$
and
$\mathcal{L}_{\text{black}}=\E\lVert\widehat{\Bsp}'-\Bsp'^{\star}\rVert_2^2$
the separately weighted latent blocks (the white block is supervised
\emph{directly}, exploiting the grey box),
$\mathcal{L}_{\mathrm{eq}}=\E\lVert\mathrm{MLP}_{\mathrm{eq}}(c)-\Lat'^{\star}_M\rVert_2^2$
training the anchor network, and ${\mathcal{L}_{\Sigma
    n_z}}=\E(\log\sum_i n_{z,i}(\widehat{\wsp}_z))^2$ the
{charge-state analogue of the logarithmic penalty $\mathcal{L}_{\Sigma
    N}$ of \Cref{sec:ae}, with $n_z$ recovered from $\widehat{\wsp}_z$
  by inverting \eqref{eq:scalenz}; the decoder network is not run
  during training, so this penalty is the sole conservation signal
  seen by the latent dynamics, while the decode-time projection
  guarantees the constraint at inference}. Every training window starts
from a true latent state, which bounds the error accumulated within a window
but does not remove the mismatch between windowed training and free-running
inference; the growing horizon and the full-rollout validation, which
integrates each validation trajectory end to end from its initial state, are
what address that mismatch.

Each expectation $\E$ in \Cref{eq:nodeloss} is a plain mean over every
scalar entry it contains: over the windows in the batch, the $H$ target points
of each window (the initial point is excluded), and the coordinates of the
block, four for $\mathcal{L}_{\text{white}}$ and six for
$\mathcal{L}_{\text{black}}$. Equal weights $\lambda_w=\lambda_b$ therefore
weight the two blocks equally per coordinate, not per block.
$\mathcal{L}_{\mathrm{eq}}$ is the mean over the batch and the ten coordinates
of the squared difference between the anchor and the terminal latent variable of
the window's own trajectory, and $\mathcal{L}_{\Sigma n_z}$ the mean over the
batch and the $H$ points. Each epoch draws four windows per training trajectory
with the start index uniform in $\{0,\dots,n_\tau-1-H\}$, in batches of $1024$
windows ($46$ optimizer steps per epoch); the horizon is
$H=2+2\lfloor e/100\rfloor$ at epoch $e$, capped at $64$, and had reached $56$
when training ended. Gradients are clipped to unit norm and all Stage-2
arithmetic is single precision. The unrolled RK4 graph for a batch of $1024$
windows of at most $64$ steps and the resident latent dataset
($14{,}800\times100\times10$ single-precision values, $59$\,MB) are small;
peak memory was not logged. The reported run took about four hours on one
NVIDIA A100 GPU for $2760$ epochs, about five seconds per epoch.

\paragraph{On the anchor network.}
{The steady-state estimate enters the damping through a small anchor network
$\Leq(c)=\mathrm{MLP}_{\mathrm{eq}}(c)$, trained against the measured
terminal latent variables through $\mathcal{L}_{\mathrm{eq}}$; the network, rather than
the measured terminal latent variables themselves, is used throughout training and inference,
because the measured terminal states exist only for the training trajectories
whereas the network supplies a smooth $\Leq(c)$ over the condition space. The six black
coordinates of $\Leq$ have no counterpart in a charge-state balance and are
known only through the encoder applied to simulated terminal states, and where
the terminal state at fixed $c$ depends on the initial condition the regression
returns an average over initial conditions.
Because the
damping is a soft pull whose per-coordinate strength is learned, the rollouts
stay accurate even when $\Leq$ is only approximate; the supporting training
diagnostics are reported in \Cref{sec:setup}.}
We call $\Leq$ an equilibrium anchor rather than an attractor because
attraction is not established. The linear form of the damping is prescribed,
and both $\bm{A}$ and $\Leq$ are fitted through the trajectory loss, not
identified physical relaxation rates. At the anchor the integrated field
equals $\Fth^{\mathrm{raw}}(\Leq,\tau;c)$, which is not constrained to vanish,
and the Jacobian of the integrated field is
$\partial_{\Lat'}\Fth^{\mathrm{raw}}-\mathrm{diag}(\bm{A})$, whose stability
positive $\bm{A}$ alone does not guarantee; the nonlinear network can also
compensate for changes in damping strength or anchor location, so the
decomposition is not uniquely identified by trajectory data. What the results
show is that rollouts reach the observed terminal states at the end of the
trained clock interval over the tested set (\Cref{sec:res-cons}). The checks
of \Cref{sec:res-field} show more: the field does not vanish at the terminal
state or at the anchor, its Jacobian there has an eigenvalue with positive
real part on almost every test trajectory, and rollouts continued past the
clock interval leave the terminal state. The anchor is therefore not a stable
fixed point of the learned field, and, with the no-damping ablation of
\Cref{sec:res-deploy}, the damping is not what brings the rollouts to the
terminal state.

\subsection{Dataset and structural properties}
\label{sec:analysis}

The dataset comprises $14{,}800$ trajectories on a grid of $10$ background
densities $\nA$, $37$ electron temperatures $\Te$, and $40$ initial conditions.
At each $(\nA,\Te)$ the $40$ initial conditions lie on a straight line in
population space between two extreme members (an affine fit leaves residuals
of order $10^{-16}$), and each is labelled by a scalar $n_{\mathrm{per}}$
stored with the trajectory, which takes the same $40$ values at every
$(\nA,\Te)$; we use $n_{\mathrm{per}}$ to identify initial conditions in the
figures and in the error breakdowns.
Each trajectory is obtained by integrating \eqref{eq:cr} from $t=10^{-16}$ to
$1\,$s and records $\Ne(t)$, together with the four charge-state partial sums
$n_z$ (group sums of the density-normalized populations $\Ntilde=\Ne/\nA$), the
radiative power loss $\Rate(t)$, and the conditions $c$. The solution is saved at
$1019$ times, $1000$ logarithmically spaced up to $0.1\,$s followed by $19$
uniformly spaced up to $1\,$s; we use the logarithmically spaced part throughout.
These are recorded output times of the implicit reference solver, which takes
its own error-controlled internal steps between them; the arc-length grid of
\Cref{sec:sampling} selects among the recorded output times and never sees the
solver's internal steps. The data are split by trajectory into
train/validation/test ($80/10/10$) at random. Because each $(\nA,\Te)$ grid
point carries $40$ trajectories, every grid point is represented in the
training set, and a test trajectory shares its conditions with about $32$
training trajectories that differ in the initial condition. The test set
therefore measures generalization across initial conditions at the trained
conditions; generalization to unseen $(\nA,\Te)$ is tested separately, with a
split that holds out whole $(\nA,\Te)$ grid points, in \Cref{sec:res-deploy}.

\begin{figure}[t]
\centering
\includegraphics[width=\linewidth]{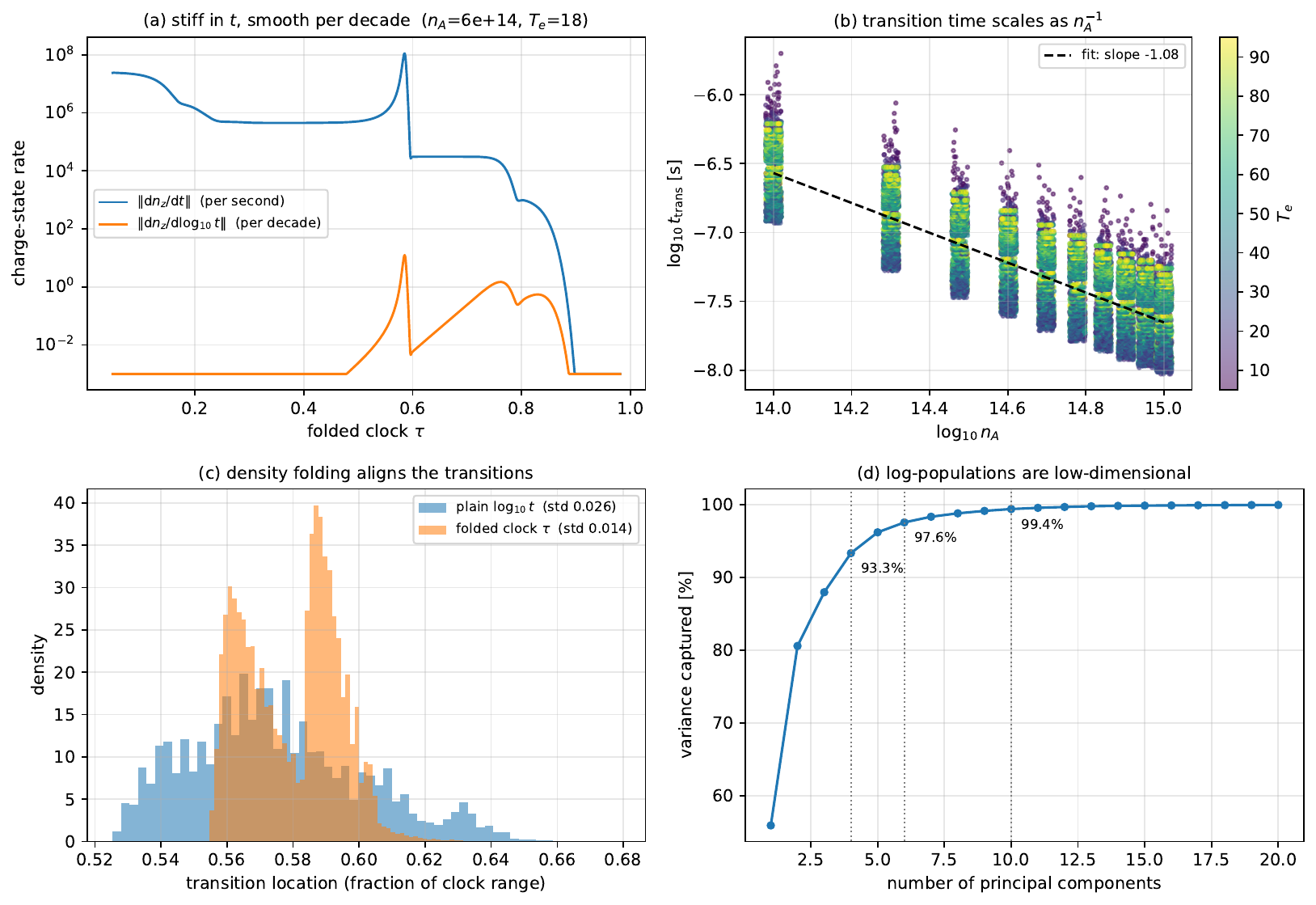}
\caption{Data properties that motivate the surrogate design, computed over all
$14{,}800$ trajectories. (a)~Charge-state rate along a representative
trajectory, per second and per decade of $\log_{10}t$: stiff in $t$, smooth in
logarithmic time. (b)~Time at which the neutral fraction falls below one half,
scaling as $t_{\mathrm{trans}}\propto n_A^{-1}$ (dashed: least-squares fit at the
median $T_e$), with negligible residual $T_e$ dependence. (c)~Spread of
transition locations before and after folding the density into the clock.
(d)~Cumulative variance of the principal components of the log-populations: four
carry the bulk and ten capture essentially all resolvable structure, motivating
the $4{+}6$ latent split in the grey-box autoencoder.}
\label{fig:data}
\end{figure}

Each design choice above exploits a measurable property of these trajectories. We quantify those properties here over the full set
(\Cref{fig:data}) and, for each property, name the design decision it supports. They fall into two groups: those that govern the treatment
of \emph{time}, and those that concern the structure of the \emph{state
space}.

The kinetics are stiff in physical time but the sampled trajectories are smooth in logarithmic time. On the
saved grid the step size ranges over
$\Delta t_{\max}/\Delta t_{\min}\approx1.3\times10^{16}$, and the ionization
transition unfolds between $10^{-8}$ and $10^{-6}\,\mathrm{s}$ within a
$1\,\mathrm{s}$ horizon. That fast feature is six to eight orders of magnitude
narrower than the integration interval, and the fast modes of \eqref{eq:cr}
are faster still, which is why an implicit solver was used to generate the data. Measured per decade of $\log_{10}t$ rather than per
second, the peak charge-state rate drops from about $10^{8}$ to about $10$
(\Cref{fig:data}a), and the transition spreads over a median $17\%$ of the clock
range. These are properties of the sampled trajectory, not of the stability
spectrum: the clock leaves the eigenvalue ratios of \eqref{eq:cr} unchanged
(\Cref{eq:alpha}). What they show is that on the clock the latent path is a
slow evolution with one localized feature, so that a modest number of steps
placed by the arc-length sampling below can follow it, which is what the
explicit integration of the learned field exploits (\Cref{sec:integrator}).

The transition also moves with density. The time at which the neutral fraction
drops below one half (the first recorded time with $n_0<1/2$, which defines
$t_{\mathrm{trans}}$ throughout; every trajectory in the dataset crosses, so
none is excluded) spans $2.3$ decades across the dataset and scales as
$\nA^{-1.08}$, with negligible residual dependence on $\Te$ (\Cref{fig:data}b).
Folding $\nA$ into the clock removes the density trend and halves the spread of
transition locations, from a standard deviation of $0.026$ to $0.014$ of the
clock range (\Cref{fig:data}c, \Cref{sec:clock}). The clock alone does not widen the
transition enough: for the sharpest trajectories it still spans only $0.3\%$ of
the $\tau$ range. On a $100$-point grid, uniform sampling puts a median of $17$
points inside the transition and none at all for the sharpest, whereas
arc-length sampling puts a median of $32$ there and never fewer than $5$
(\Cref{sec:sampling}).

Three properties of the state space shape the representation. First, the
charge fractions are approximately stationary by the end of almost every
trajectory. The charge-state drift over the final $50$ samples has median
$2\times10^{-8}$, and the dominant terminal fractions are set by the conditions:
across the dataset they move from singly ionized at low $\Te$ to fully stripped at
high $\Te$, with a smooth crossover. At fixed conditions, trace populations such
as the post-transition $n_0$ settle to values that depend on the initial
condition (\Cref{fig:charge-nper}) and remain constant to $1\,$s. The target of
the damping term can therefore be estimated from the conditions only up to
that spread (\Cref{sec:vf}). The spread of the drift is wide: over all
$14{,}800$ trajectories its $95$th percentile is $2.7\times10^{-5}$, and $2\%$
of the trajectories drift by more than $10^{-3}$, all at $\Te\le10$\,eV and
$\nA\le5\times10^{14}$; the largest, $9.2\times10^{-2}$, occurs at
$\nA=10^{14}$ and $\Te=7.5$\,eV, where the charge fractions are still evolving
at the final saved time. At $\Te\ge20$\,eV the drift is below $2.2\times10^{-6}$
for every trajectory. Approximate stationarity of the four charge fractions
therefore holds at the end of most, not all, trajectories, and it does not
establish convergence of every level or latent coordinate, uniqueness of the
terminal state, or attraction toward it; the accuracy of the surrogate at the
final time is reported in \Cref{sec:res-deploy}. Second, the populations
compress well. Principal components of the log-populations capture $93\%$ of the
variance in four, $98\%$ in six and $99.4\%$ in ten (\Cref{fig:data}d), so ten
latent coordinates, four fixed to the charge states and six learned, retain
essentially all resolvable variance (\Cref{sec:aearch}). Captured
variance supports compressibility; whether the $4{+}6$ split is sufficient, and
whether the low-variance channels are dynamically relevant, is established only
by the reconstruction and rollout errors of \Cref{sec:results}. Third, the radiative
power loss is smooth in the populations. Its logarithm is approximately linear in
the $94$ log-populations ($R^2\approx0.97$) over roughly five decades, so it
decodes from the same latent variable under the standardization of
\Cref{sec:ae}.

\section{Numerical Results}
\label{sec:results}

We evaluate the surrogate (the frozen grey-box autoencoder of \Cref{sec:ae}
composed with the latent neural ODE of \Cref{sec:node}) on the
held-out test trajectories. Unless stated otherwise, every rollout is produced by
integrating the latent dynamics from the \emph{true initial latent variable only} and
decoding to physical charge states $n_z$ and radiative power loss $\Rate$; no
ground-truth information enters the rollout after $\tau_0$. We report the
training behaviour (\Cref{sec:setup}), charge-state rollout fidelity and its
distribution over the parameter space (\Cref{sec:res-rollout}), the radiative
power loss (\Cref{sec:res-rpl}), mass conservation, grid refinement, and
long-rollout behaviour (\Cref{sec:res-cons}), deployable grids, unseen
conditions, integration error, and ablations (\Cref{sec:res-deploy}), the
learned field near the terminal state, restarts, and perturbations
(\Cref{sec:res-field}), and a controlled comparison with
the discrete flow-map surrogate (\Cref{sec:res-fmnet}).

\subsection{Experimental setup and training behaviour}
\label{sec:setup}

\Cref{tab:hparams} summarizes the architecture and the principal training
settings. The autoencoder maps the $94$-level scaled population vector to the
ten-dimensional grey-box latent space (four physically fixed white coordinates
plus six
learned black coordinates) and is trained once; the latent neural ODE is a small
model
trained on the frozen latent variables.

\begin{table}[t]
\centering
\caption{Architecture and principal training settings of the reported models.}
\label{tab:hparams}
\begin{tabular}{@{}lll@{}}
\toprule
& Stage 1 (autoencoder) & Stage 2 (latent neural ODE) \\
\midrule
latent / state dim     & $94\to 10$ ($4$ white $+\,6$ black) & $10$ \\
encoder / vector field & MLP $(64,32,32)$ & MLP $(128,128,128,128)$ \\
decoder                & MLP $(32,32,64)$ & --- \\
activation             & SiLU & SiLU \\
clock embedding        & --- & $K=4$ Fourier features \\
anchor network         & --- & MLP $(64,64)$ \\
time sampling          & --- & arc-length adaptive ($n_\tau=100$) \\
integrator             & --- & explicit RK4 \\
conservation           & log penalty $+$ simplex proj. & log penalty $+$ decode-time simplex proj. \\
{log floors $\varepsilon_N{=}\varepsilon_R$} & {$10^{-30}$} & {$10^{-30}$} \\
rollout curriculum     & --- & $H:2\to 56$ \\
optimizer / LR         & Adam, $10^{-3}$ (cosine) & Adam, $10^{-3}$ (plateau $\times0.5$/$100$ ep.) \\
batch size             & $8192$ & $1024$ \\
loss weights           & $w_W{=}1$, $w_R{=}1.5$, $w_m{=}1$ & $\lambda_w{=}\lambda_b{=}\lambda_{\mathrm{eq}}{=}1$, $\lambda_c{=}0.1$ \\
epochs                 & $3000$ & $2760$ (best epoch $2740$) \\
precision              & mixed (AMP) & single (fp32) \\
windows / grad.\ clip   & --- & $4$ per traj.\ \& epoch / norm $1$ \\
\bottomrule
\end{tabular}
\end{table}

\begin{figure}[t]
\centering
\includegraphics[width=\linewidth]{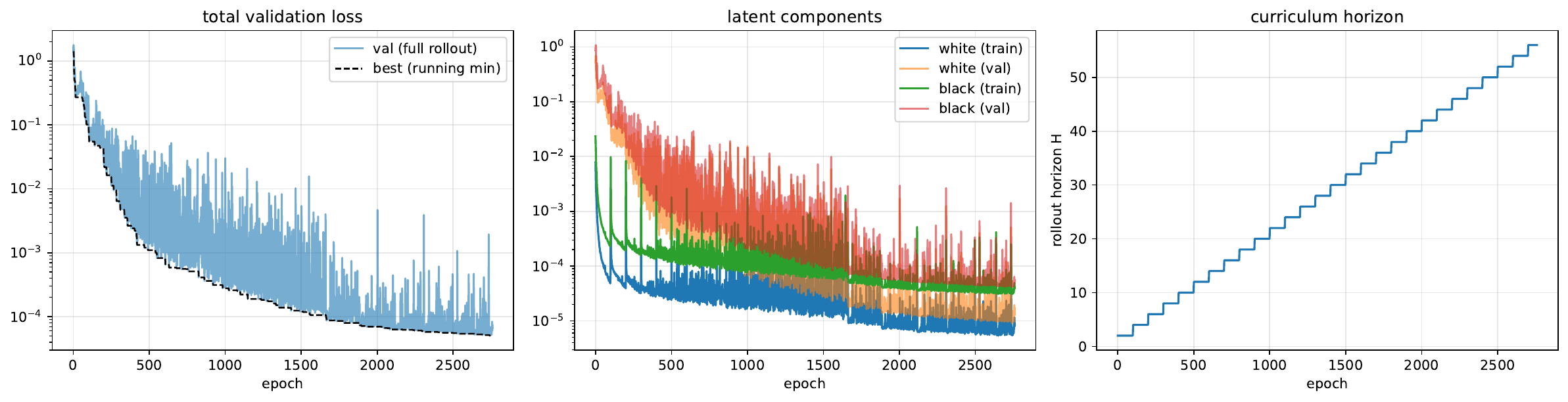}
\caption{Stage-2 training. \emph{Left:} full-rollout validation loss (per epoch
and running minimum), settling near $5\times10^{-5}$. \emph{Center:} latent loss
split into white (charge-state) and black (learned) blocks for train and
validation; the directly supervised white block converges fastest and lowest.
\emph{Right:} the curriculum rollout horizon $H$, increased in stages from $2$
to ${\sim}\,56$.}
\label{fig:training}
\end{figure}

\Cref{fig:training} shows the Stage-2 optimization. The full-rollout validation
loss, obtained by integrating each validation trajectory end to end from its
true initial latent variable and measuring the latent MSE, decreases by more than
four
orders of magnitude, and its running minimum settles near $5\times10^{-5}$. The
per-block curves show the grey box behaving as intended: the white
(charge-state) coordinates, which are supervised directly, converge fastest and
to the lowest error (${\sim}\,10^{-5}$), while the six learned black coordinates
settle near $10^{-4}$. {The anchor loss plateaus near
$1.2\times10^{-2}$, an RMS error in $\Leq$ of roughly $0.1$ per coordinate
in standardized latent units, while the full-rollout validation error continues
to fall: the rollouts do not require an accurate $\Leq$, although whether
this reflects the softness of the pull or compensation by the nonlinear field
is not separated here.} The different loss levels of the two blocks show
that the learned coordinates are harder to fit than the directly supervised
ones; they do not by themselves establish that the black block carries dynamics
independent of the charge states. The right panel shows the curriculum: the rollout
horizon $H$ is increased in stages from $2$ to ${\sim}\,56$ steps over training,
so the model first learns one-step accuracy and is then progressively exposed to
longer horizons; every window starts from a true latent state, and the
full-rollout validation loss, computed on the $1480$ validation trajectories
after every epoch, selects the reported checkpoint (the epoch of its minimum,
$2740$) and drives both the learning-rate decay and the early-stopping rule of
$300$ epochs without improvement, which was not triggered: the run ended at
epoch $2760$ with the horizon at $56$ of a possible $64$.

\paragraph{Inference cost.}
The trained surrogate is fast enough for in-loop use: the $58{,}526$-parameter
model (the vector field, $53{,}514$ parameters; the anchor network,
$5{,}002$; and the ten damping strengths; the frozen decoder is not counted)
completes all $1480$ full-trajectory test rollouts ($99$ RK4 steps each,
on the $100$-point arc-length grid) in $113$\,ms on a single GPU, i.e.\
$76\,\mu$s per full trajectory or ${\sim}1.3\times10^{4}$ trajectories per
second. The timing covers the latent rollout only (vector field, damping,
and anchor, in single precision, on one NVIDIA A100 GPU of NERSC Perlmutter):
it is the wall-clock time of the batched rollout call, averaged over ten
repetitions after one warm-up call, with device synchronization before and
after; the frozen decoder, the encoding of the initial state, host--device
transfers, and the construction of the grid (precomputed and cached) are not
included, and the run-to-run variability was not recorded. The figure is an
amortized throughput, all $1480$ trajectories being integrated as one batch,
and not the latency of a single call. The $374\,\mu$s per trajectory of
\Cref{tab:fmnet} was measured the same way on batches of $300$ (five
repetitions): $300\times374\,\mu$s is $112$\,ms, the same time per batch
call as $1480\times76\,\mu$s, because the call is dominated by the $396$
sequential small-kernel evaluations of the field (four per RK4 step) rather
than by arithmetic, so the per-trajectory cost falls inversely with the batch
size until the GPU saturates. The step count makes a
qualitative point: $99$ steps of an \emph{explicit} integrator suffice for a
full trajectory of the learned latent field, whereas an explicit integrator
advancing \eqref{eq:cr} itself in physical time would have its step bounded by
the fastest mode present at each instant; we have not measured that step count,
and the reference data were generated with an implicit solver.

\paragraph{Stage-1 reconstruction baseline.}
Because the neural ODE predicts the latent variable and the frozen decoder
produces the
physical
state, the autoencoder's own reconstruction error is a natural baseline for
the full pipeline (not a strict lower bound, since a predicted latent state can
occasionally decode closer to the truth than the encoded reference does). \Cref{tab:ae-err} reports it over the complete test set: the mean
absolute error on the charge fractions is $1.6\times10^{-3}$, the radiative loss
is reconstructed to a median $2.2\times10^{-3}$ in $\log_{10}$, and the simplex
projection holds $|\sum_i n_{z,i}-1|$ below $4\times10^{-7}$ everywhere. The full
pipeline's rollout MAE of $4.2\times10^{-3}$ (\Cref{tab:rollout-err}) is within a
factor of $2.6$ of this baseline, so the error added by the learned
dynamics is of the same order as that of the representation itself. The two
numbers are not computed alike: the reconstruction MAE is a mean over all
$1480\times1000$ native-grid samples of the test set, whereas the rollout MAE
is the median over test trajectories of a per-trajectory mean on the
$100$-point arc-length grid, which weights the transition more heavily; the
factor is indicative, not a decomposition of the error.

\begin{table}[t]
\centering
\caption{Stage-1 autoencoder reconstruction over the full test set
($1480$ trajectories $\times\,1000$ time points).}
\label{tab:ae-err}
\begin{tabular}{@{}lcccc@{}}
\toprule
 & $n_0$ & $n_1$ & $n_2$ & $n_3$ \\
\midrule
rel.\ $L_2$ & $1.18\times10^{-2}$ & $3.07\times10^{-2}$ & $3.21\times10^{-2}$ & $9.34\times10^{-3}$ \\
\midrule
\multicolumn{5}{@{}l}{MAE ($n_z$) $=1.61\times10^{-3}$;\quad RMSE ($n_z$) $=7.80\times10^{-3}$;\quad
RPL $|\Delta\log_{10}|$ median $=2.2\times10^{-3}$}\\
\multicolumn{5}{@{}l}{$|\sum_i \Ntilde_i-1|$: mean $=4.5\times10^{-8}$, max $=3.6\times10^{-7}$}\\
\bottomrule
\end{tabular}
\end{table}
\subsection{Charge-state rollout fidelity across the parameter space}
\label{sec:res-rollout}

\begin{figure}[h]
\centering
\includegraphics[width=\linewidth]{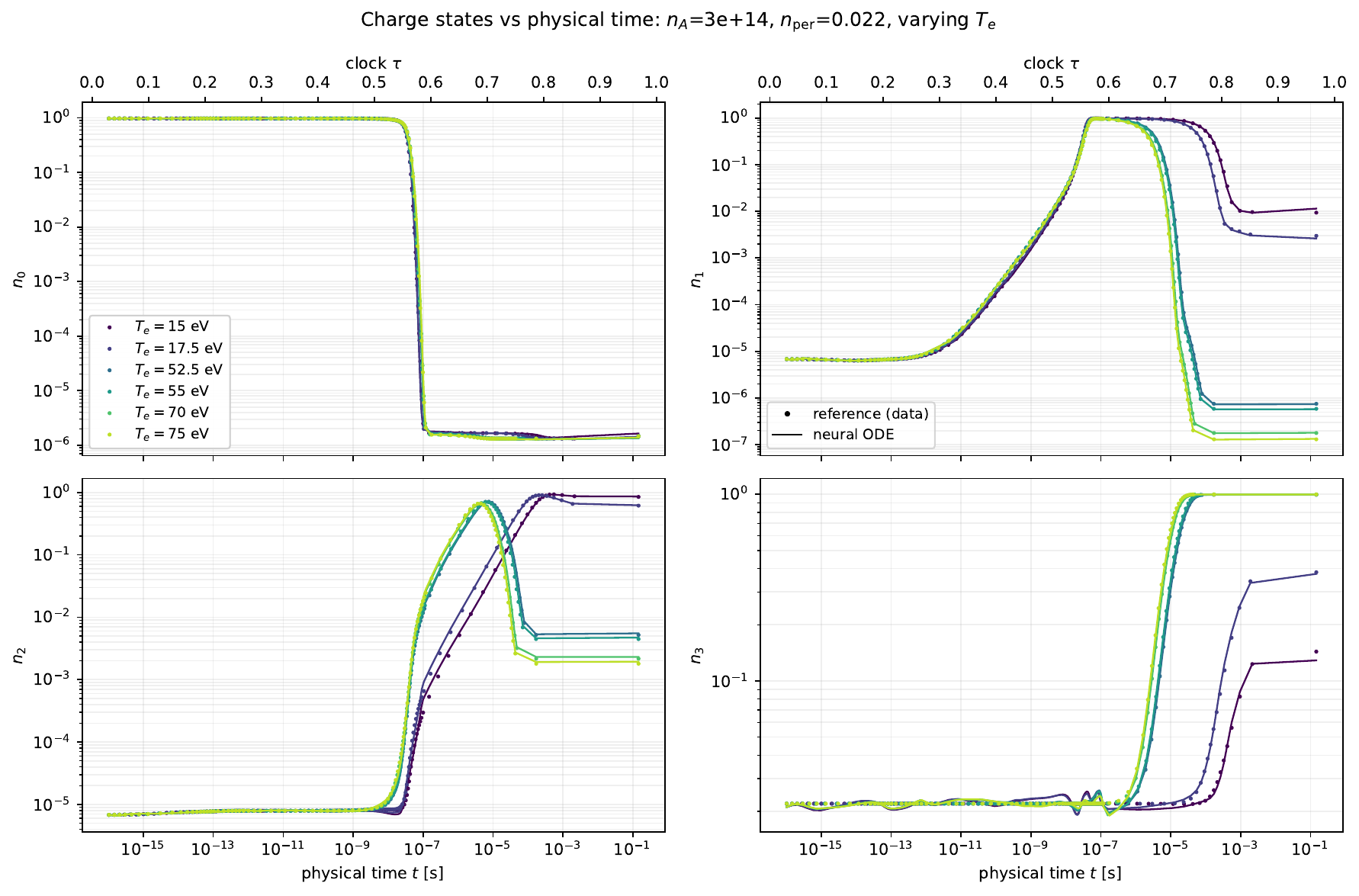}
\caption{Charge-state rollouts versus CR data at fixed
$\nA=3\times10^{14}$ and a fixed initial condition, sweeping $\Te$. Panels are the
four lithium stages $n_0$ (neutral) to $n_3$ (fully stripped); dots are the CR
ground truth and lines the decoded neural ODE rollout from the true initial
condition.
Each colour is one test trajectory; bottom axis: physical time $t$, top
axis: clock $\tau$, as in \Cref{fig:charge-nper}. The terminal
ionization moves from $n_1$-dominated at low $\Te$ to $n_3$-dominated at high
$\Te$, reproducing the terminal state of the data, with the largest
visible departure in $n_3$ at the final time for $\Te=15$\,eV.}
\label{fig:charge-Te}
\end{figure}

\begin{figure}[h]
\centering
\includegraphics[width=\linewidth]{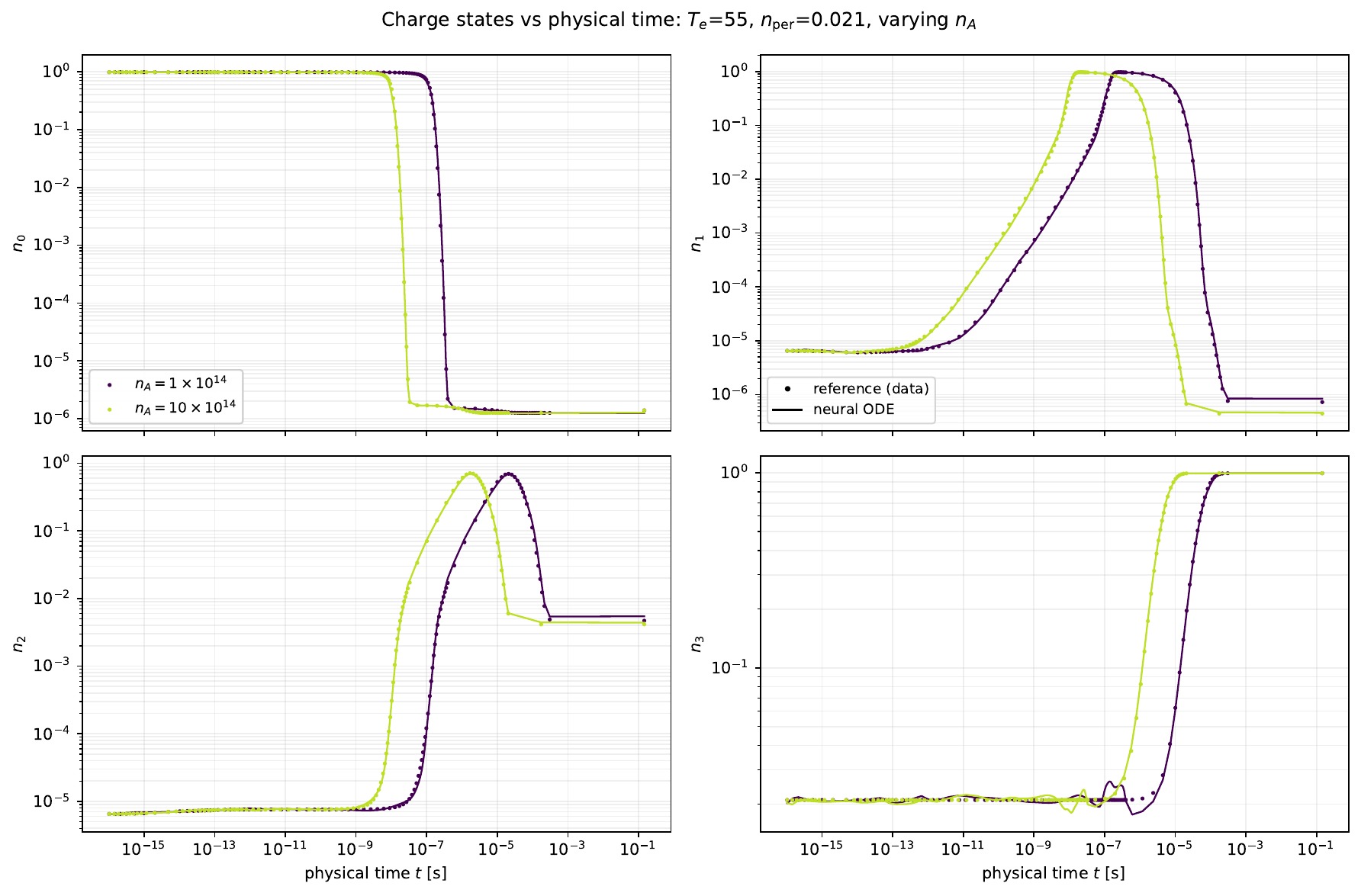}
\caption{Charge-state rollouts versus CR data at fixed $\Te=55$,
a fixed initial condition, varying $\nA$. Horizontal axis: physical time
$t$; the clock depends on $\nA$, so no clock axis is shown, and the shift of
the transition with $\nA$ is read directly in $t$.}
\label{fig:charge-nA}
\end{figure}

\begin{figure}[h]
\centering
\includegraphics[width=\linewidth]{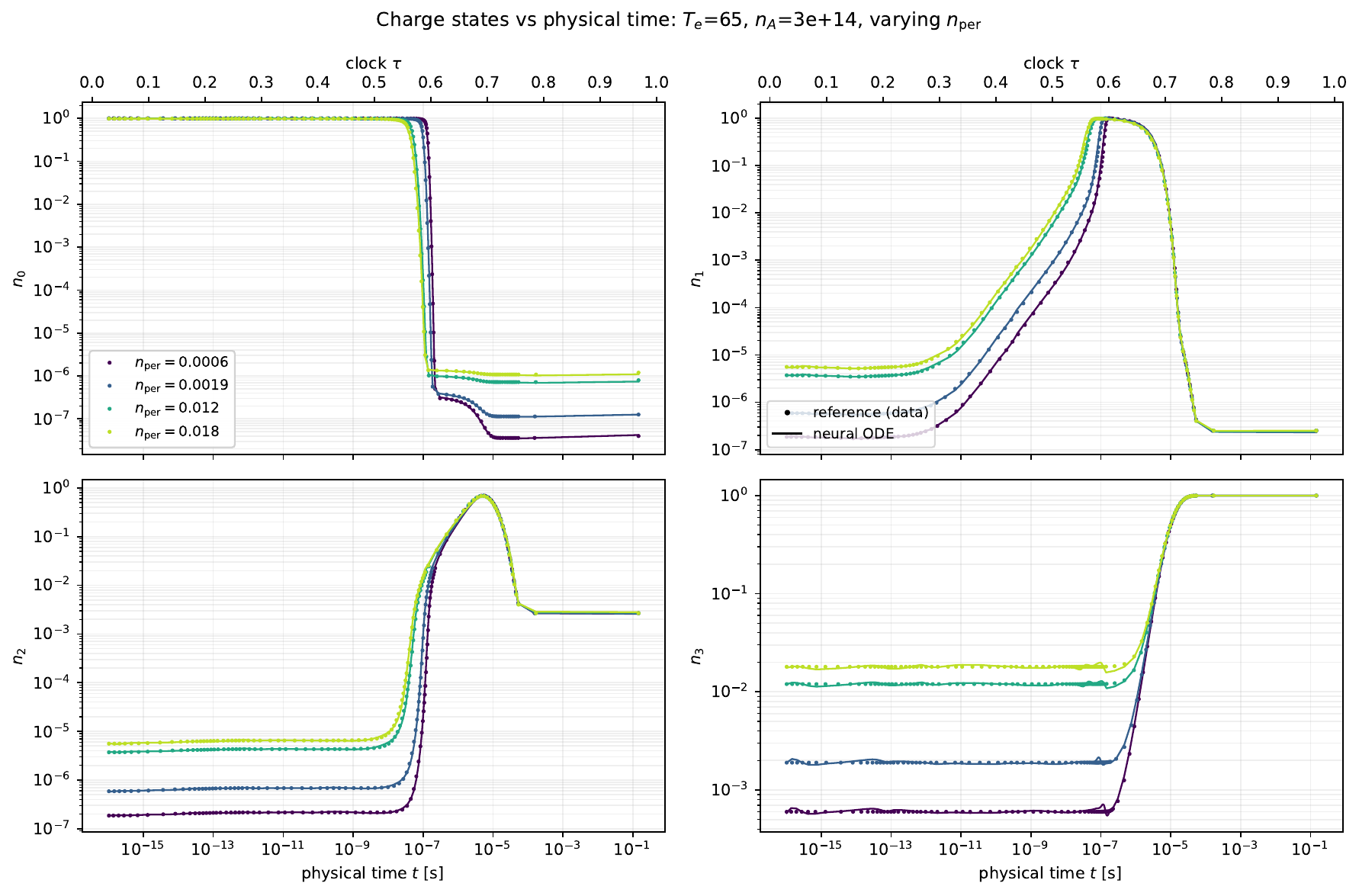}
\caption{Charge-state rollouts versus CR data at fixed $\Te=65$,
$\nA=3\times10^{14}$, varying the initial-condition parameter
the initial condition. The pre- and post-transition floors span more than seven
decades; the surrogate follows them, with only sub-percent oscillation in the
$n_3$ floor. Bottom axis: physical time $t$; top axis: clock $\tau$.}
\label{fig:charge-nper}
\end{figure}

Figures~\ref{fig:charge-Te}--\ref{fig:charge-nper} show decoded full-trajectory
rollouts (solid lines) against the ground-truth CR data (dots) for the four
lithium charge states $n_0$ (neutral) through $n_3$ (fully stripped), sweeping one
condition at a time and holding the other two fixed. In each panel the
test trajectories are overlaid against physical time $t$ on a logarithmic
axis; the dots are the data at the points of each trajectory's $100$-point
arc-length grid, and when $\nA$ is fixed a top axis gives the clock $\tau$.
 Throughout this section the plotted
and scored charge fractions are the decoded partial sums of the projected
populations, not the white latent coordinates predicted by the neural ODE. The agreement spans the
entire dynamic range of the problem, six to seven decades, including the deep
post-transition floor of the neutral fraction and the small high-ionization
channels.

Varying $\Te$ at fixed $\nA$ and initial condition (\Cref{fig:charge-Te}) exercises
the steady-state endpoint: the terminal ionization shifts smoothly from
$n_1$-dominated at low $\Te$ to $n_3$-dominated at high $\Te$, the
$\Te$ dependence of the CR steady state, and the surrogate
reproduces both the transition location and the terminal state; this is
endpoint accuracy over the tested trajectories, not independent evidence about
the equilibrium of \eqref{eq:cr}. Varying $\nA$
(\Cref{fig:charge-nA}) shifts the transition in time by the expected
density-linear amount ($t_{\mathrm{trans}}\propto\nA^{-1}$); because the integration clock folds in
$\nA$, the surrogate tracks this shift without retraining. Varying the
initial condition (\Cref{fig:charge-nper}) changes
the starting distribution and the post-transition floors over more than seven
decades, which the surrogate also follows. The only visible departures are
oscillations in the pre-transition plateau of $n_3$, the smallest
channel in that regime at $10^{-3}$ to $2\times10^{-2}$ of the total; they are
largest just before the transition, with dips of up to about a fifth of the
plateau value, a few times $10^{-3}$ in absolute fraction (\Cref{fig:charge-nA}),
and small differences at the final saved time of the lowest-$\Te$
trajectories (\Cref{fig:charge-Te}). They do not shift the transition.

\begin{figure}[t]
\centering
\includegraphics[width=0.72\linewidth]{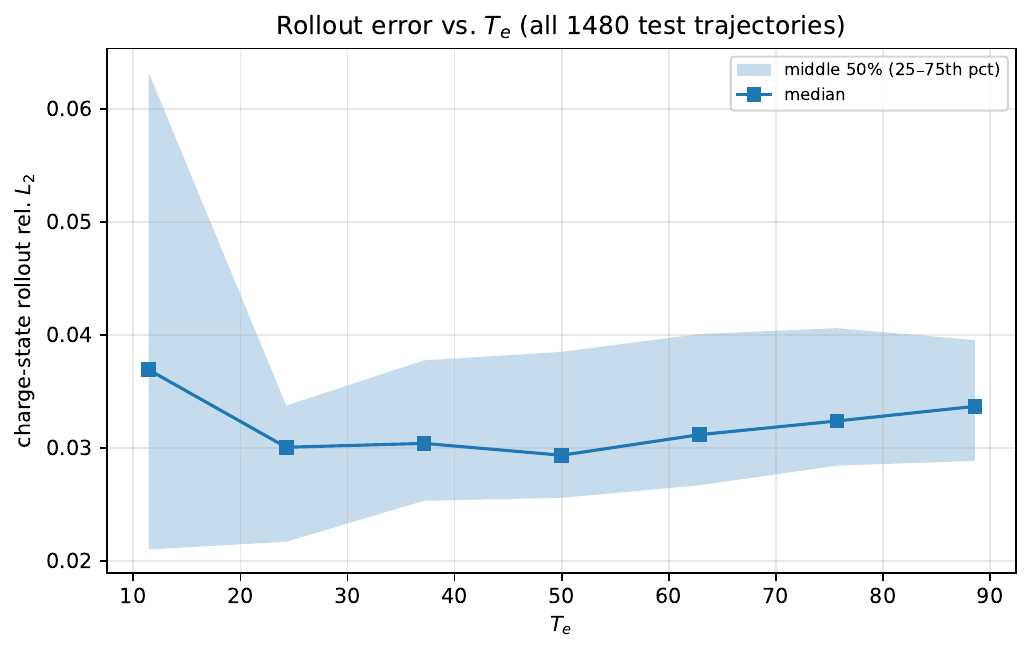}
\caption{Charge-state rollout relative $L_2$ error over all $1480$ test
trajectories, binned by $\Te$ (median and interquartile band). The median is
${\approx}\,3\%$ and nearly flat in $\Te$; the spread widens only at the lowest
temperatures, where the transition is sharpest.}
\label{fig:rollout-err}
\end{figure}

To quantify accuracy over the \emph{entire} test set rather than selected
trajectories, \Cref{fig:rollout-err} reports the per-trajectory charge-state
rollout relative $L_2$ error, comparing the decoded $n_z(t)$ to the CR data over
the full trajectory, for all $1480$ held-out trajectories, binned by $\Te$. The median error is ${\approx}\,3\%$ and is nearly
flat across the temperature range, with the central $50\%$ of trajectories between
roughly $2\%$ and $4\%$. The spread widens only at the lowest temperatures, where
the transition is sharpest and the post-transition populations smallest. The
accuracy seen in the example trajectories is therefore representative of the whole
parameter grid. The relative and absolute measures are complementary and are defined as
follows. For test trajectory $i$ on its own $100$-point arc-length grid, let
$\hat n_{z}(\tau_k)$ be the decoded charge fractions (partial sums of the
projected decoded populations over the level groups $1$--$31$, $32$--$62$,
$63$--$93$, and $94$) and $n_{z}(\tau_k)$ the reference charge fractions,
obtained from the white block of the reference latent on the same grid through
\eqref{eq:scalenz} and normalized to unit sum. With $\lVert\cdot\rVert$ the
Euclidean norm over the $100$ grid points, equally weighted, the relative error
is
$e_i=\sum_z\lVert\hat n_z-n_z\rVert\big/\big(\sum_z\lVert n_z\rVert+10^{-30}\big)$,
the channels being summed before the ratio is taken, so a channel that is near
zero over much of the trajectory contributes little to the denominator and
cannot inflate $e_i$; the MAE and RMSE of trajectory $i$ are the mean and
root-mean-square of $|\hat n_z-n_z|$ over the $100$ points and four channels.
\Cref{tab:rollout-err} reports the median and maximum of each over the $1480$
test trajectories. All three are grid-weighted: the arc-length grid places
about a third of its points in the transition, so they emphasize the
transition relative to a physical-time or uniform-clock quadrature; a
uniform-clock score, on the solver's saved output times, is given in
\Cref{sec:res-deploy}. The relative error normalizes by the size of the signal and
is the stricter measure for trajectories whose fractions are small over much of
the clock; the absolute error on fractions in $[0,1]$ that sum to one reflects
accuracy on the dominant populations.

\begin{table}[t]
\centering
\caption{Charge-state rollout error over the $1480$ held-out test trajectories,
as the channel-aggregated relative $L_2$ and as the absolute error on the
charge fractions $n_z\in[0,1]$ (definitions in the text). Each entry is the
median or maximum over trajectories of a per-trajectory score on the $100$-point
arc-length grid; the conservation column is over all $1480\times100$ decoded
states.}
\label{tab:rollout-err}
\begin{tabular}{@{}lcccc@{}}
\toprule
& rel.\ $L_2$ (channel-aggregated) & MAE ($n_z$) & RMSE ($n_z$) & $|\sum_i n_{z,i}-1|$ \\
\midrule
median (test) & $3.1\times10^{-2}$ & $4.2\times10^{-3}$ & $1.5\times10^{-2}$ & $0$ \\
max (test)    & $1.3\times10^{-1}$ & $1.3\times10^{-2}$ & $6.3\times10^{-2}$ & $3.6\times10^{-7}$ \\
\bottomrule
\end{tabular}
\end{table}

\subsection{Radiative power loss}
\label{sec:res-rpl}
\begin{figure}[t]
\centering
\includegraphics[width=\linewidth]{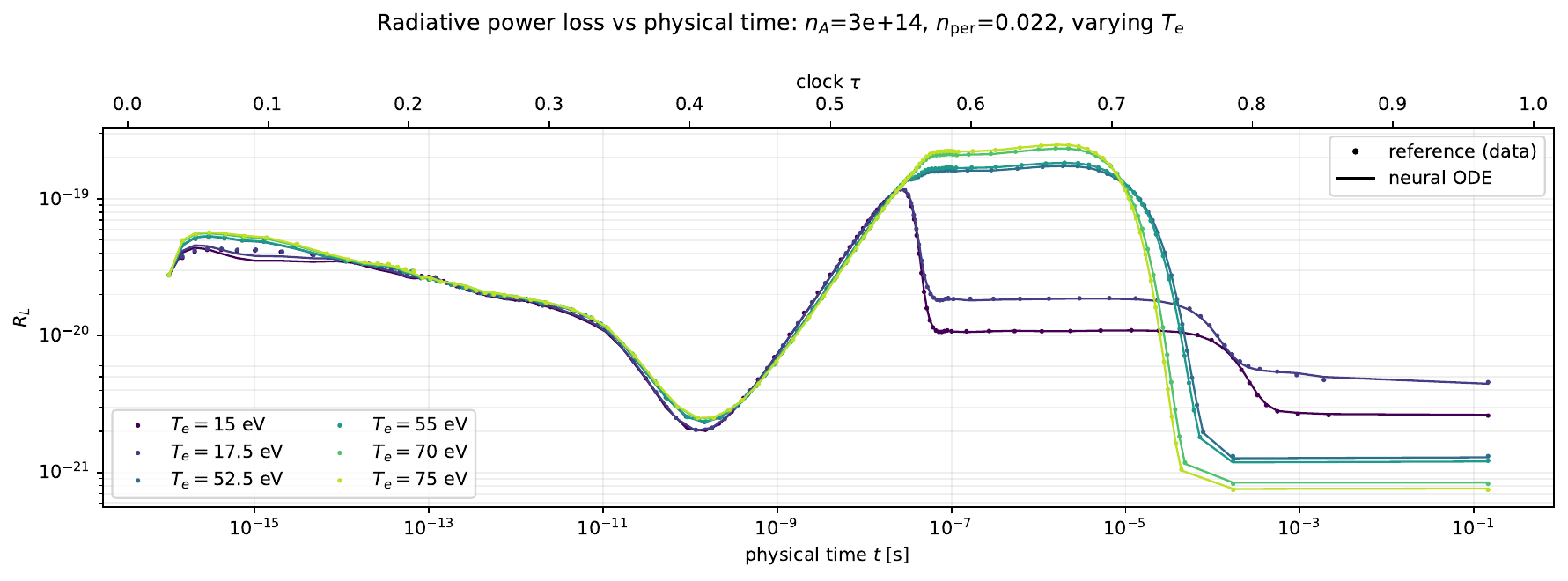}
\caption{Radiative power loss $\Rate(t)$ versus CR data at fixed
$\nA=3\times10^{14}$ and a fixed initial condition, sweeping $\Te$ (dots: truth;
lines: neural ODE). The non-monotonic rise--dip--peak structure and its $\Te$
dependence are reproduced over two to three decades. Bottom axis:
physical time $t$; top axis: clock $\tau$.}
\label{fig:rpl-Te}
\end{figure}

\begin{figure}[h]
\centering
\includegraphics[width=\linewidth]{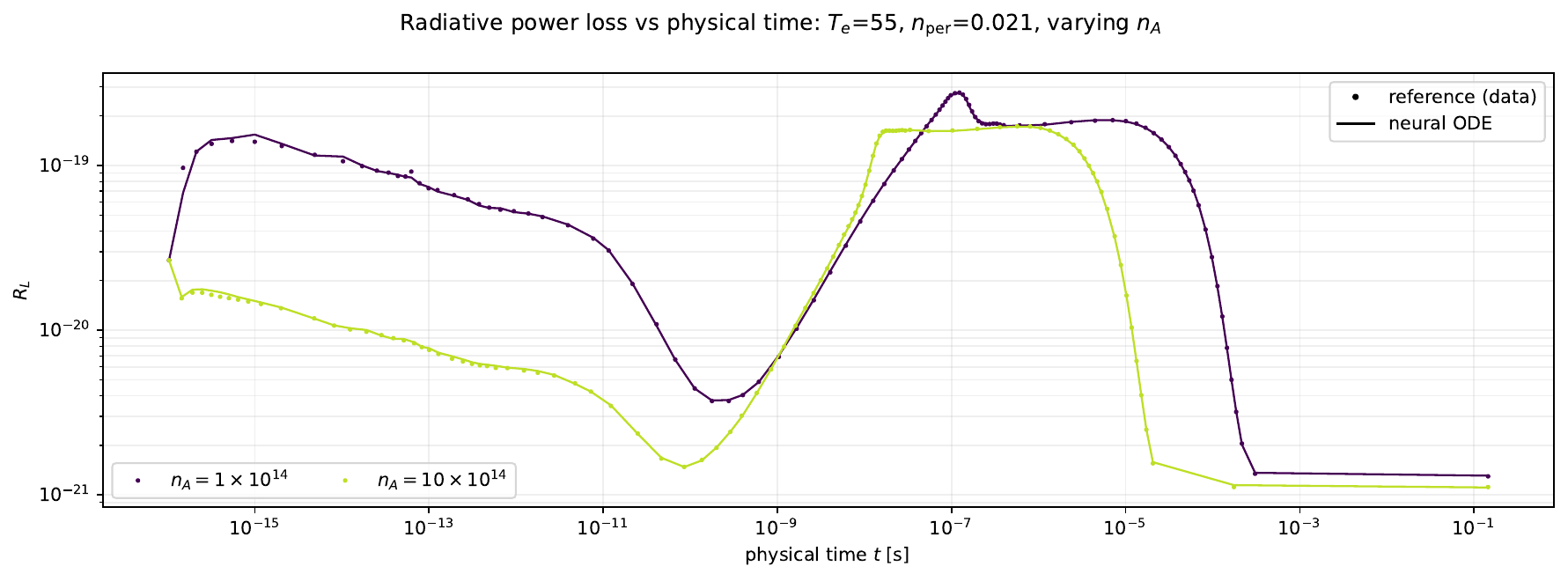}
\caption{Radiative power loss at fixed $\Te=55$ and a fixed initial condition,
varying $\nA$, against physical time $t$.}
\label{fig:rpl-nA}
\end{figure}

\begin{figure}[h]
\centering
\includegraphics[width=\linewidth]{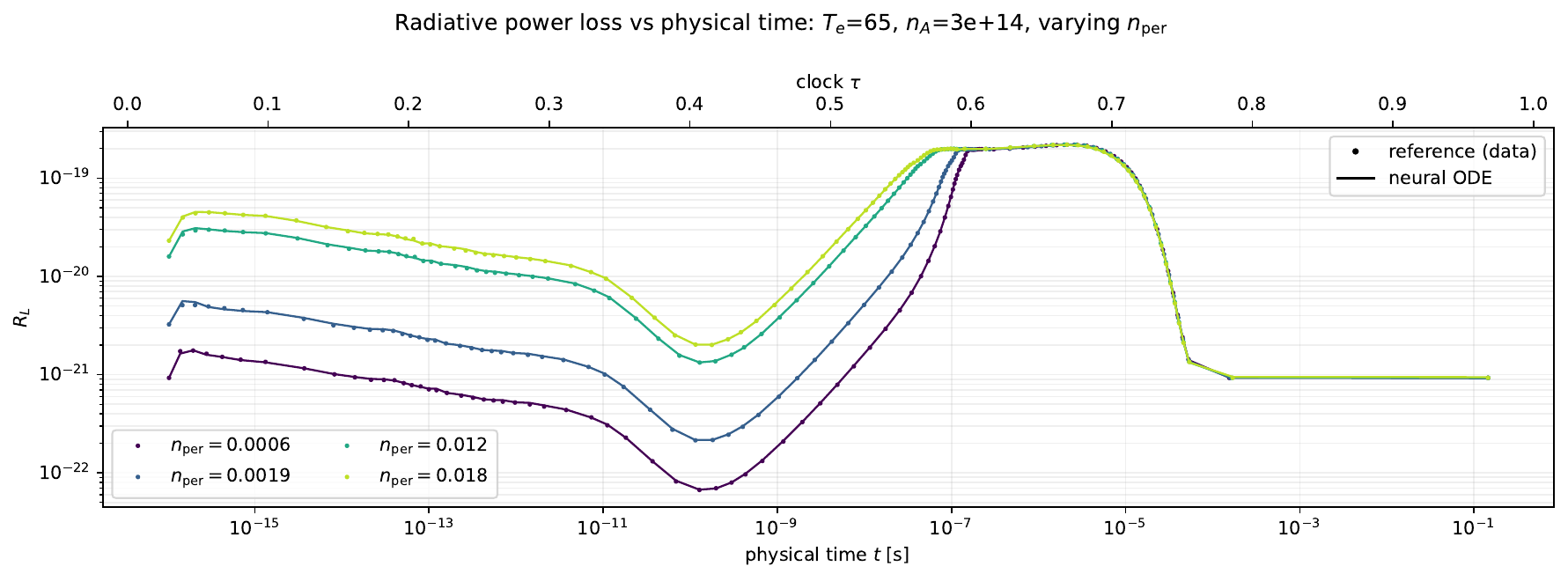}
\caption{Radiative power loss at fixed $\Te=65$, $\nA=3\times10^{14}$, varying
the initial condition; bottom axis: physical time $t$, top axis: clock
$\tau$.}
\label{fig:rpl-nper}
\end{figure}

The radiative power loss $\Rate$ is the quantity most directly consumed by an
integrated transport code, and it is decoded from the same latent variable as the
populations, exploiting the near-linear dependence of the log radiative loss on
the log-populations. Neither the encoder nor the decoder receives
$(\nA,\Te)$; the conditions never enter Stage~1. The decoder is a map from the
ten latent coordinates alone, and the condition dependence of the decoded
populations and of $\Rate$ is carried by the latent state itself: the four
white coordinates fix the charge-state fractions, and the six black
coordinates, fitted pointwise to population vectors sampled over all
conditions, encode the within-charge-state level distributions, which at any
instant are determined by the populations rather than labelled by $c$.
$\Rate$ is the solver's radiative-loss output at each saved time, standardized
as $\log_{10}\Rate$ with a single global mean and standard deviation over the
training set (\Cref{sec:ae}). We report $\Rate$ in the solver's output units; every error
in $\Rate$ below is given in decades of $\log_{10}\Rate$ or as a relative
error, so none depends on the choice of units. Figures~\ref{fig:rpl-Te}--\ref{fig:rpl-nper} show
$\Rate(t)$ for the same parameter sweeps. Unlike the monotone charge-state
transitions, $\Rate$ is strongly non-monotonic (an early rise, a sharp dip as
the neutral stage burns through, and a broad recombination/line-emission peak
before the final decay) and varies over two to three decades. The surrogate
reproduces this full structure, including the depth and location of the dip and
the height of the main peak, across all three sweeps. $\Rate$ is decoded
from the latent state rather than modelled dynamically, so no separate
dynamical model is required for it; its rollout error therefore has two
sources, the reconstruction error of the decoder (median $2.2\times10^{-3}$ in
$\log_{10}$ over the test set, \Cref{tab:ae-err}) and the latent rollout error,
which a sensitive radiation decoder can amplify. Over the full test set, the
error $|\Delta\log_{10}\Rate|$ averaged over each trajectory's grid has median
$9.9\times10^{-3}$, $95$th percentile $2.5\times10^{-2}$ and maximum
$5.3\times10^{-2}$; its largest single-point value per trajectory has median
$0.23$ and maximum $0.78$ decades. The two sources separate as follows:
comparing the decoded reference latent with the data gives a
trajectory-averaged error of median $8.9\times10^{-3}$, and comparing the
decoded rollout with the decoded reference latent gives $4.1\times10^{-3}$, so
the decoder, not the latent dynamics, dominates the radiation error. (These
trajectory averages on the arc-length grid weight the transition and are larger
than the pointwise median of \Cref{tab:ae-err}.) Against the data at the
$1000$ saved output times, the transition maximum of $\Rate$ (the maximum after
the early dip) has height error median $2.4\times10^{-3}$ decades ($95$th
percentile $2.2\times10^{-2}$, maximum $7.0\times10^{-2}$), with the largest
medians at $\Te\le7.5$\,eV, and the time at which $\Rate$ first reaches half of that
maximum is reproduced to median $0$, $95$th percentile $0.015$, and maximum
$0.06$ decades of physical time; the time of the maximum itself is less well
defined on the flat plateau (median $0.015$, maximum $1.9$ decades). The
radiated energy $\int\Rate\,\diff t$, integrated in physical time up to the end
of the charge-state transition, has relative error median $0.4\%$, $95$th
percentile $2.7\%$, and maximum $12\%$, largest at low $\Te$; over the whole
window, which the terminal plateau dominates, the median is $1.4\%$ and the
maximum $23\%$. The early maximum, within the first decade of physical time, is
the least accurate feature: its height error has median $0.18$ and $95$th
percentile $0.56$ decades, and its time error median $0.05$ decades. The latent
error also peaks there (\Cref{fig:grid-invariance}); whether the decoder, the
rollout, or the interpolation between grid points dominates this error has not
been separated.

\subsection{Mass conservation and long-rollout stability}
\label{sec:res-cons}

Three separate properties are evaluated separately here. First, mass
conservation is enforced \emph{by construction}: {the decoder normalizes
the level-resolved output onto the simplex (\eqref{eq:simplex}), and the
charge-state fractions, as its partial sums, inherit} $\sum_i n_{z,i}=1$ to the projection
tolerance at every output time: measured over the full test rollout,
$|\sum_i n_{z,i}-1|$ has median $6\times10^{-8}$ and maximum $6\times10^{-7}$
(\Cref{tab:rollout-err}). The predicted populations are therefore always a valid
distribution, in contrast to the soft penalty of the prior flow-map surrogate,
under which the sum could drift. This guarantee concerns the decoded sum
only: it says nothing about the white latent block, whose predicted charge
fractions are a second estimate of $n_z$ that need not coincide with the decoded
partial sums, and nothing about kinetic accuracy, which the rollout errors above
measure. Over the test set the two estimates differ: the white fractions are
the more accurate pointwise (MAE median $1.1\times10^{-3}$, against
$4.2\times10^{-3}$ for the decoded partial sums), but they are not
conservative (the largest $|\sum_z n_z-1|$ along a trajectory has median
$1.4\times10^{-2}$ and maximum $0.19$), and the largest difference between
the two estimates along a trajectory has median $0.11$ and maximum $0.43$.
Throughout this paper the decoded partial sums are the quantity scored and
plotted.

\begin{figure}[t]
\centering
\includegraphics[width=.72\linewidth]{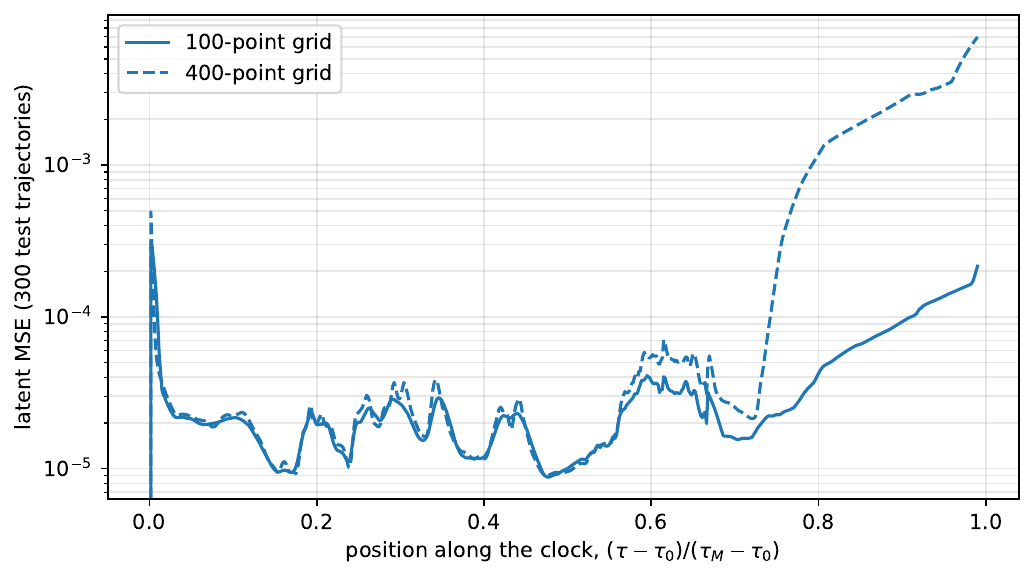}
\caption{Per-step latent error of the full model along the test rollout on the
$100$-point (solid) and $400$-point (dashed) output grids, plotted against
position along the clock, $(\tau-\tau_0)/(\tau_M-\tau_0)$; each
trajectory's error is interpolated to common clock positions before averaging
over the first $300$ test trajectories. Up to about $70\%$ of the clock,
through the initial transient and the ionization transition, the two curves
nearly coincide. On the late plateau beyond about $75\%$ they separate: the
error grows on both grids, to about $2\times10^{-4}$ ($100$ points) and
$7\times10^{-3}$ ($400$ points), so the learned field is not grid-converged
there. The final $1\%$ of the clock is not shown. At the final time the decoded
charge fractions on the $100$-point grid have absolute error of median
$2.2\times10^{-4}$ and maximum $9.8\times10^{-2}$ over the full test set
(\Cref{sec:res-deploy}).}
\label{fig:grid-invariance}
\end{figure}

Second, the charge-state error of the rollout is robust to refinement of
its grid, but the latent error on the late plateau is not. Because the model is a continuous-time integrator, the output grid is
not fixed at training. All grids used here are arc-length grids built from the
reference trajectory (\Cref{sec:sampling}), and the output samples and the
integration steps coincide: there is no internal substepping. Rolling the
same trained model out on arc-length grids of $100$, $200$, and $400$ points
leaves the test error unchanged (MAE $4.5$, $4.6$, and $4.6\times10^{-3}$
respectively, on the first $200$ test trajectories). Over the first
$70\%$ of the clock, which contains the initial transient and the ionization
transition, the per-step latent error is nearly the same on the $100$- and
$400$-point grids (\Cref{fig:grid-invariance}): there it is a function of
\emph{position along the trajectory} rather than of the number of steps
integrated. On the late plateau it is not. Beyond about $75\%$ of the clock the
latent error grows on both grids, and faster on the finer one, to about
$2\times10^{-4}$ on the $100$-point grid and $7\times10^{-3}$ on the
$400$-point grid at $99\%$ of the clock. The learned field, trained through the
$100$-point discretization, is therefore not grid-converged on the late
plateau, consistent with the comparison against an error-controlled
integration in \Cref{sec:res-deploy}. The grid-weighted charge-state MAE above
does not show this because the plateau carries little weight on arc-length
grids. On such a grid the
surrogate can
therefore be queried at any temporal resolution at or above that of training;
below it the explicit rollout eventually under-resolves the transition (on a
$50$-point grid the latent rollout becomes non-finite), which sets one deployment constraint. The
scores in this section are computed on the arc-length grid of each row and are
therefore grid-weighted; a score on a common, independently specified set of
times is given in \Cref{sec:res-deploy}. The same
trained model rolled out on a uniform $100$-point grid scores \emph{lower}, MAE
$3.1\times10^{-3}$ and median relative error $1.9\times10^{-2}$ on the same
$200$ trajectories, not because it is more accurate there but because the
uniform grid places most of its points on the plateaus, where the error is
small; this is the grid weighting made visible.

Third, long rollouts are empirically robust within the trained horizon: every rollout in
Figures~\ref{fig:charge-Te}--\ref{fig:charge-nper} is integrated to the end of the
horizon and reaches the observed terminal distribution (final-time
error median $2.2\times10^{-4}$, \Cref{sec:res-deploy}). This is robustness on
the tested cases within the clock range of the data, not dynamical stability:
\Cref{sec:res-field} shows that the terminal state is not attracting for the
learned field, that perturbations can grow transiently, and that rollouts
continued past the clock range leave the terminal state; a model trained
without the damping term is as accurate at the median
(\Cref{sec:res-deploy}).

\subsection{Deployable grids and ablations}
\label{sec:res-deploy}
This subsection addresses three questions that the protocol above leaves
open: whether the accuracy survives when the integration grid is not built
from the reference trajectory, whether it survives at conditions $(\nA,\Te)$
absent from training, and how much of the error the fixed-step RK4 integration
contributes. It also reports ablations of three components of Stage~2 and the
error distributions and tails over the full test set.

\paragraph{Common scoring.} The scores of \Cref{tab:rollout-err} are taken on
each trajectory's own grid and depend on where that grid places its points. To
compare grids, every rollout is here also interpolated linearly in $\tau$ to
the solver's $1000$ saved output times, which are uniformly spaced in $\tau$,
and its decoded charge fractions are scored there against the CR data; we call
the result the common-grid MAE. On the reference arc-length grid its median is
$1.6\times10^{-3}$, against $4.2\times10^{-3}$ on the arc-length grid itself,
because the arc-length grid places about a third of its points in the
transition, where the error is largest.

\paragraph{Grids that do not use the reference trajectory.} Each grid is given
as fractions of the trajectory's clock interval $[\tau_0,\tau_M]$, which
follows from $\nA$ and the output window and is therefore known at deployment.
\Cref{tab:grids} compares, all with $100$ points: the reference arc-length grid
of \Cref{sec:sampling}; a uniform grid; the arc-length grid averaged over all
training trajectories, over the training trajectories at the same $\nA$, or
over those at the nearest training condition; and a two-pass grid, for which
the model is rolled out on the training-averaged grid, an arc-length grid is
built from its own predicted latent path by \eqref{eq:arclen}, and the model
is rolled out again on that grid. The two-pass grid uses only the initial
state, the conditions, and the trained model.

At the median every grid is within a factor of $1.5$ of the reference grid;
the grids differ in the tails. The single-pass grids fixed in advance have
$95$th percentiles three to seven times, and maxima six to thirteen times,
those of the reference grid, on trajectories whose transition does not fall
where the fixed grid is dense. On some of these trajectories the latent rollout
leaves the physical range: on the training-averaged grid the white charge
fractions of the worst trajectory sum to $3\times10^{5}$, while the decoded
fractions, normalized by \eqref{eq:simplex}, still form a valid distribution
with a bounded error. The simplex normalization can therefore hide a failed
latent rollout, which the white block reveals. The two-pass grid removes these
failures: its median, $95$th percentile, and maximum are close to those of the
reference grid, and the largest deviation of its white fractions from unit sum
is $2.9$. It costs two rollouts and one grid construction, $153\,\mu$s per
trajectory against $76\,\mu$s, measured as in \Cref{sec:setup}. It is thus a
deployable replacement for the reference grid with nearly the same accuracy.

\begin{table}[t]
\centering
\caption{Accuracy of the main-text model on grids that do (reference) and do
not use the reference trajectory, over the $1480$ test trajectories: common-grid
MAE of the decoded charge fractions, scored at the solver's $1000$ saved
output times. Time per trajectory as in \Cref{sec:setup}; the two-pass time
includes both rollouts and the construction of the second grid.}
\label{tab:grids}
\begin{tabular}{@{}lccccc@{}}
\toprule
grid & RK4 steps & median & $95$th pct. & max & $\mu$s/traj. \\
\midrule
reference arc length & $99$ & $1.6\times10^{-3}$ & $2.8\times10^{-3}$ & $5.3\times10^{-3}$ & $76$ \\
uniform & $99$ & $2.3\times10^{-3}$ & $7.9\times10^{-3}$ & $3.0\times10^{-2}$ & $76$ \\
training mean, all & $99$ & $1.8\times10^{-3}$ & $2.0\times10^{-2}$ & $4.9\times10^{-2}$ & $76$ \\
training mean, same $\nA$ & $99$ & $1.7\times10^{-3}$ & $1.9\times10^{-2}$ & $5.5\times10^{-2}$ & $77$ \\
training mean, nearest $(\nA,\Te)$ & $99$ & $1.7\times10^{-3}$ & $8.7\times10^{-3}$ & $7.0\times10^{-2}$ & $76$ \\
two-pass (model's own path) & $198$ & $1.6\times10^{-3}$ & $4.2\times10^{-3}$ & $7.7\times10^{-3}$ & $153$ \\
\bottomrule
\end{tabular}
\end{table}

\begin{figure}[t]
\centering
\includegraphics[width=\linewidth]{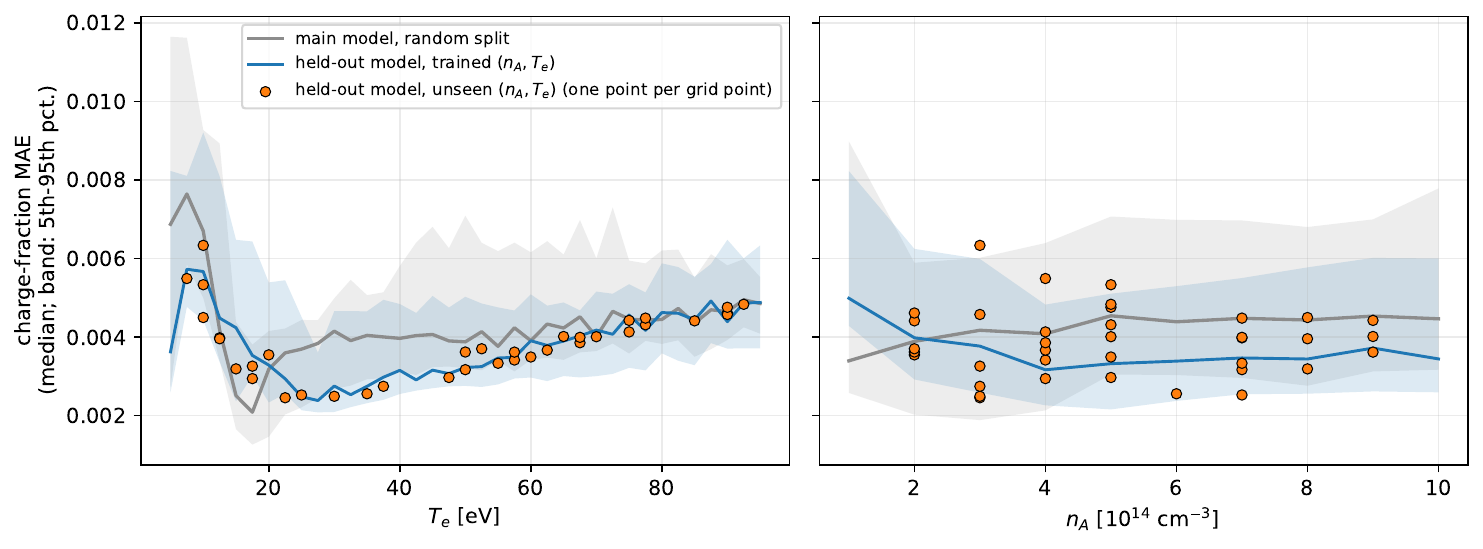}
\caption{Median per-trajectory MAE of the decoded charge fractions on the
reference grid, against $\Te$ (left) and $\nA$ (right). Grey: model of the main
text, random-split test set; blue: held-out model at trained grid points
($1184$ trajectories); bands: $5$th to $95$th percentile over the trajectories
at each value. Orange dots: held-out model at each of the $37$ unseen grid
points, median over its $40$ initial conditions.}
\label{fig:heldout-err}
\end{figure}

\begin{figure}[t]
\centering
\includegraphics[width=\linewidth]{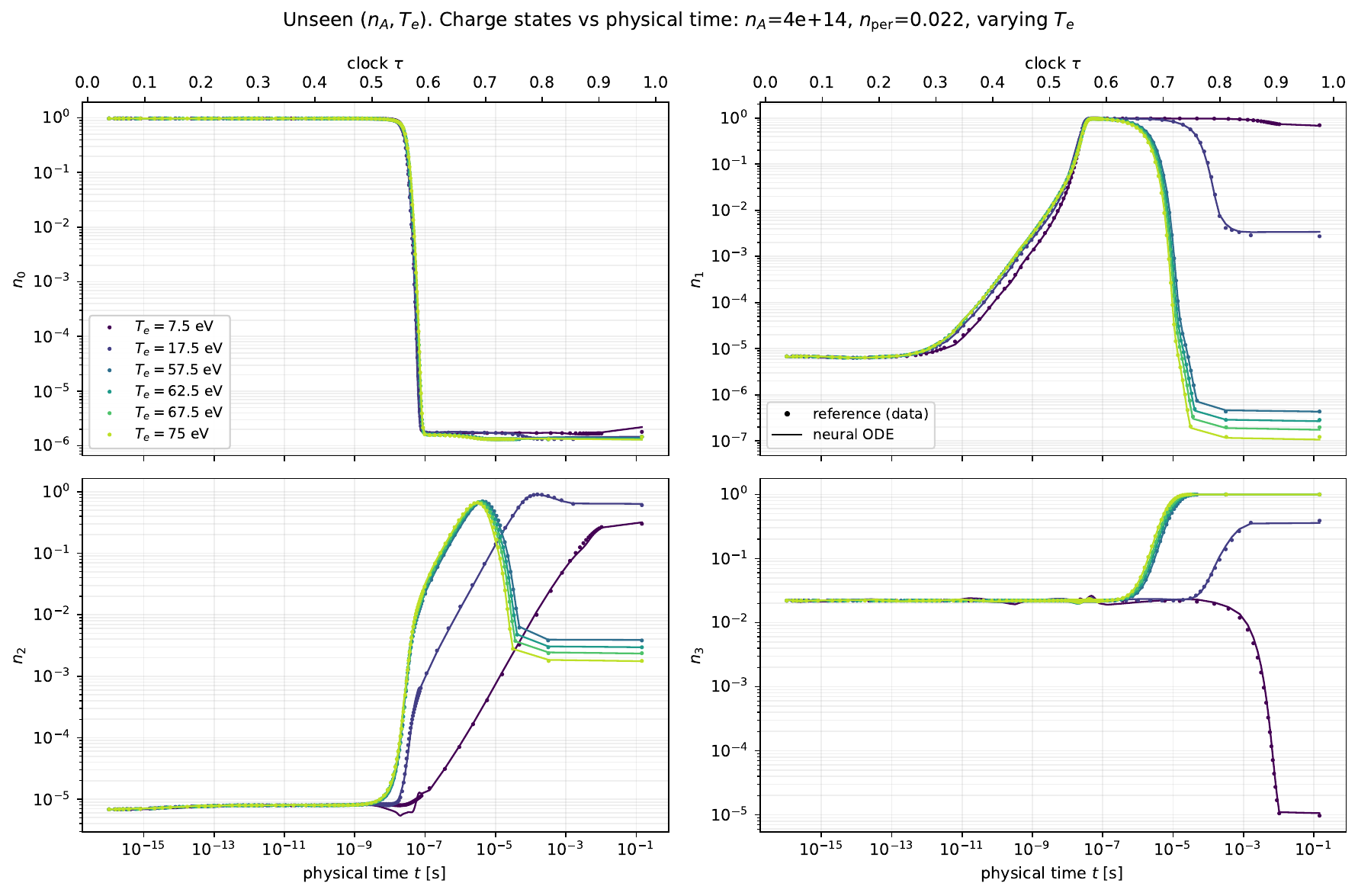}
\caption{Charge-state rollouts of the held-out model at unseen
conditions: $\nA=4\times10^{14}$, $n_{\mathrm{per}}=0.022$, and the six held-out
values of $\Te$ at this density ($7.5$, $17.5$, $57.5$, $62.5$, $67.5$, and
$75$\,eV), none of which has a training trajectory at this $\nA$. Dots: CR data
at the points of each trajectory's arc-length grid; lines: decoded neural ODE
rollout from the true initial condition. Bottom axis: physical time $t$; top
axis: clock $\tau$. The rollouts follow the data through the transition and
onto the terminal state for every $\Te$, including the lowest, $7.5$\,eV, where
$n_2$ and $n_3$ evolve on a different time scale from the others.}
\label{fig:heldout-Te}
\end{figure}

\begin{figure}[t]
\centering
\includegraphics[width=\linewidth]{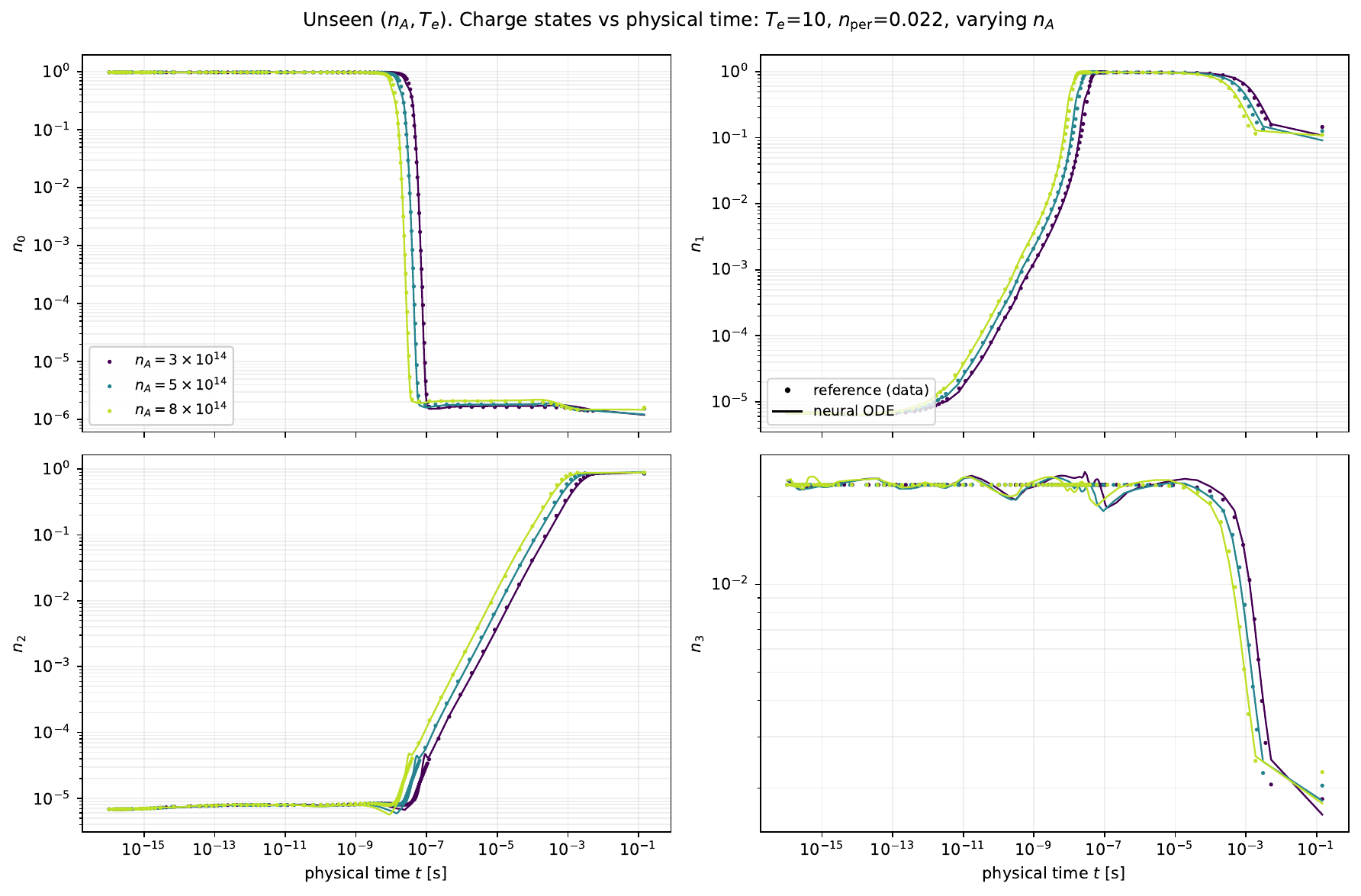}
\caption{As \Cref{fig:heldout-Te}, at $\Te=10$\,eV and the three held-out
densities at this temperature, $\nA=3$, $5$, and $8\times10^{14}$. The clock
depends on $\nA$, so only physical time is shown. The transitions shift to
earlier times with increasing $\nA$ and are reproduced; the visible departures
are oscillations of the small $n_3$ plateau before the transition and
differences in $n_1$ and $n_3$ around $10^{-3}$\,s.}
\label{fig:heldout-nA}
\end{figure}

\paragraph{Unseen conditions.} The split of \Cref{sec:analysis} places
every $(\nA,\Te)$ grid point in training. To test generalization to conditions,
whole grid points are held out: $37$ of the $370$ points ($1480$ trajectories,
all $40$ initial conditions at each) form the test set and another $37$ the
validation set, drawn at random from the interior of the grid, excluding the
extreme values of $\nA$ and $\Te$, so that the test is interpolation between
trained conditions. Of the trajectories at the remaining $296$ points, $10\%$
($1184$) are also withheld from training, to score the same model at trained
conditions and new initial conditions. Both stages are retrained from scratch
on the remaining $10{,}656$ trajectories with the settings of
\Cref{tab:hparams} ($3000$ epochs for Stage~1, with the gradient norm clipped
at $1$; $3350$ for Stage~2, against $2760$ for the main-text model), and the
Stage-2 latents and their standardization are recomputed with the new
autoencoder. A first attempt without the clipping diverged at epoch $395$ of
Stage~1: one decoded $\log_{10}$ population left the range of the data, its
exponential overflowed in the conservation penalty, and every later step was
non-finite. The retrained autoencoder therefore also caps the decoded
$\log_{10}$ populations at $2$ inside that penalty, a bound that a trained
model never reaches; no non-finite value occurred in the retraining, nor in the
training of the main-text model. \Cref{tab:heldout} shows no loss of accuracy at unseen
conditions: the median MAE is $3.8\times10^{-3}$ there and
$3.7\times10^{-3}$ at trained conditions for the same model, both close to
the $4.2\times10^{-3}$ of the main-text model on its random split, and the
largest error at unseen conditions, $6.6\times10^{-3}$, is below the largest
at trained ones. As on the random split, the error is largest at the lowest
temperatures (median $5.5$--$5.6\times10^{-3}$ at the held-out points with
$\Te=7.5$ and $10$). \Cref{fig:heldout-err} resolves the comparison by
condition: at each $\Te$, the median error at an unseen grid point is between
$0.75$ and $1.14$ times that of the same model at trained points with the same
$\Te$, so the dependence of the error on $\Te$ is the same at seen and unseen
conditions. \Cref{fig:heldout-Te,fig:heldout-nA} show rollouts at unseen grid
points, sweeping $\Te$ at $\nA=4\times10^{14}$ and $\nA$ at $\Te=10$\,eV, the
regime with the largest errors; they are in the format of
\Cref{fig:charge-Te}. The radiation error at unseen conditions is
$1.0\times10^{-2}$, as at trained ones. This is one random draw of held-out
points in the interior of the grid; extrapolation beyond the trained range of
$(\nA,\Te)$ is not tested.

\begin{table}[t]
\centering
\caption{Generalization to unseen conditions. Per-trajectory scores on each
trajectory's $100$-point reference arc-length grid (median, $95$th percentile,
and maximum over trajectories), and the median common-grid MAE. ``main'': the
model of the main text; ``held-out'': the model retrained on the held-out-condition split.}
\label{tab:heldout}
\small
\setlength{\tabcolsep}{4pt}
\begin{tabular}{@{}llcccccc@{}}
\toprule
model & test set & traj. & \multicolumn{3}{c}{MAE ($n_z$)} & rel.\ $L_2$ & common-grid \\
 & & & median & $95$th pct. & max & median & MAE median \\
\midrule
main & random split & $1480$ & $4.2\times10^{-3}$ & $7.2\times10^{-3}$ & $1.3\times10^{-2}$ & $3.1\times10^{-2}$ & $1.6\times10^{-3}$ \\
held-out & unseen $(\nA,\Te)$ & $1480$ & $3.8\times10^{-3}$ & $5.7\times10^{-3}$ & $6.6\times10^{-3}$ & $2.6\times10^{-2}$ & $1.3\times10^{-3}$ \\
held-out & trained $(\nA,\Te)$ & $1184$ & $3.7\times10^{-3}$ & $6.0\times10^{-3}$ & $9.8\times10^{-3}$ & $2.7\times10^{-2}$ & $1.3\times10^{-3}$ \\
\bottomrule
\end{tabular}
\end{table}

\paragraph{Integration error.} The same learned field was integrated to
the same output points with an adaptive Dormand--Prince 5(4) scheme in double
precision (relative and absolute tolerances $10^{-7}$ and $10^{-9}$). The
median MAE is then $4.3\times10^{-3}$, against $4.2\times10^{-3}$ with RK4, so
at the median the error is that of the learned field, not of its integration.
The largest difference between the RK4 and the adaptive charge fractions along
a trajectory has median $2.7\times10^{-3}$, but reaches $0.26$ on the worst
trajectory: the $100$-point RK4 rollout is not a converged solution of the
learned field everywhere. Because the field was trained through this RK4
discretization, the accuracy reported here is that of the discretized model,
and the field should not be paired with another integrator without
re-validation. The adaptive integration needs a median of $3556$ field
evaluations per trajectory, against $396$ for RK4.

\paragraph{Ablations.} Three variants of Stage~2 were each trained once with
the settings of \Cref{tab:hparams} on the same latent variables and split, each
changing one component, and evaluated on the reference grids of the $1480$
test trajectories. Without the damping term ($\bm{A}=0$ in
\eqref{eq:Fimpl}), the median MAE is $4.25\times10^{-3}$ against
$4.19\times10^{-3}$ for the full model, and the median relative $L_2$ error
$3.15\times10^{-2}$ against $3.14\times10^{-2}$; the largest relative $L_2$
error is $0.17$ against $0.13$. Trained on grids uniform in $\tau$ instead of
arc-length grids, and evaluated on the same arc-length grids as the full model,
the model is as accurate (median MAE $4.20\times10^{-3}$, largest relative
$L_2$ error $0.095$); the arc-length grid therefore matters where the rollout is
evaluated (\Cref{tab:grids}), not for training. With the scalar clock in place
of the embedding \eqref{eq:fourier}, training stopped early at a validation
loss about twelve times that of the full model, and some test rollouts were not
finite. On this dataset the damping term and the arc-length training grid
therefore give no measurable gain in accuracy and are presented as design
choices, not requirements; the clock embedding is the one ablated component
whose removal degraded the model. Each variant is a single training run, so
differences of the size seen in the largest errors may not be significant.

\paragraph{Error distributions.} For the main-text model on the reference grid,
the median MAE per $\Te$ is highest at $5$--$10$\,eV ($6.7$--$7.6\times10^{-3}$),
lowest at $15$--$17.5$\,eV ($2.1$--$2.5\times10^{-3}$), and near
$4\times10^{-3}$ above $25$\,eV. It rises mildly with density, from
$3.4\times10^{-3}$ at $\nA=10^{14}$ to $4.5\times10^{-3}$ at
$\nA\ge5\times10^{14}$, with $95$th percentiles below $9\times10^{-3}$ at every
density. Over the initial conditions it falls from $5.5\times10^{-3}$ for the
fifth with the smallest $n_{\mathrm{per}}$ to $3.9\times10^{-3}$ for the fifth
with the largest. The largest errors, up to $1.3\times10^{-2}$, occur at the
lowest density and temperatures. At the final time the absolute error of the
decoded charge fractions has median $2.2\times10^{-4}$, $95$th percentile
$1.8\times10^{-2}$, and maximum $9.8\times10^{-2}$; this is endpoint accuracy
over the tested set, not evidence of convergence to the CR steady state.

\subsection{The learned field near the terminal state and perturbations}
\label{sec:res-field}
The checks of this subsection use the main-text model on the $1480$ test
trajectories and their reference grids; none needs retraining.

\paragraph{Anchor error in physical variables.} Decoded through the frozen
decoder, the anchor $\Leq(c)$ and the measured terminal latent variable differ
in the charge fractions by a median $9\times10^{-5}$ (largest difference over
the four states; $95$th percentile $4.2\times10^{-3}$, maximum
$2.5\times10^{-2}$) and in $\Rate$ by a median $2.9\times10^{-3}$ decades
(maximum $2.2\times10^{-2}$). Against the data at the final saved time, the
decoded anchor has charge-fraction error median $1.2\times10^{-4}$ and maximum
$2.9\times10^{-2}$, close to that of the decoder alone applied to the measured
terminal latent variable ($1.4\times10^{-4}$ and $3.7\times10^{-2}$); it is
largest at $\Te=7.5$--$25$\,eV (medians up to $2\times10^{-2}$), with medians
below $10^{-3}$ from $35$\,eV up. The rollouts do not end at the anchor: in the
standardized latent space the final rollout state lies at median distance $0.03$
from the measured terminal latent variable and $0.28$ from the anchor.

\paragraph{The learned field at the terminal state.} The integrated field does
not vanish at the measured terminal latent variable: its norm there is a median
$12\%$ of the median norm along the trajectory ($2\%$ of the maximum), and the
same holds at the anchor. The Jacobian $\partial\Fth/\partial\Lat'$ of the
integrated field, damping included, has an eigenvalue with positive real part at
the measured terminal state on $99\%$ of the test trajectories (largest real
part median $0.95$, maximum $22$; most negative real part median $-7.9$), and
at two thirds of the grid points along the rollout. The terminal state is
therefore neither a fixed point nor locally attracting for the learned field;
the rollouts reach it because the trained clock interval ends there. Continued
past that interval by a further quarter of each trajectory's clock span, in
$25$ RK4 steps, the rollouts move away: the largest change of the decoded
charge fractions has median $2.9\times10^{-2}$ and $95$th percentile $0.78$,
and the white fractions sum to as much as $1.25\times10^{2}$. The surrogate must
not be integrated beyond the trained clock interval. The same Jacobians measure
the RK4 steps against the stability of the learned field: the product of step
size and largest eigenvalue modulus exceeds the real-axis stability limit
$2.785$ of classical RK4 on $88\%$ of the trajectories, on a median of one step
in a hundred (largest product per trajectory median $6.6$, maximum $22$). The
fixed-step rollout therefore leaves the linear stability region of the learned
field on a few steps, consistent with its differences from the error-controlled
integration of \Cref{sec:res-deploy}; the field was trained through these
steps.

\paragraph{Restarts and perturbations.} Restarted from the true latent state at
grid index $25$, $50$, or $75$ with that state's own clock value, the rollout
has the accuracy of the uninterrupted rollout on the remaining points (median
MAE $5.6$, $8.2$, and $5.0\times10^{-3}$ for both), so the error is not carried
by accumulated state error. Restarting from the model's own state at its clock
value reproduces the uninterrupted rollout exactly, since RK4 is a one-step
method. Restarted from the same true states with the clock reset to $\tau_0$,
the rollout fails: the decoded output is non-finite on $17$--$72\%$ of the
trajectories, depending on the restart point, and the median MAE of the rest is
$0.15$ to $0.24$. The field is not invariant under a shift of the clock, and a restart
requires the clock value of the trajectory. Random perturbations of norm $0.01$
or $0.1$ of the standardized latent state at the same three points grow
transiently, by a median factor of up to $1.4$, and by the final time have
decayed to a median $0.3$ to $0.55$ of their initial size; the $95$th
percentile of that ratio is $0.8$ to $1.2$, so on some trajectories they are not
reduced, and the largest growth is a factor of $26$.
For norm $0.1$ the final decoded charge fractions change by a median of at most
$2\times10^{-4}$ and at most $0.2$. No CR solution exists for the perturbed
states, so this measures the sensitivity of the learned field, not its
accuracy.

\subsection{Comparison with the discrete flow-map surrogate}
\label{sec:res-fmnet}
The direct predecessor of this work \citep{Xie2024} pairs the same grey-box
autoencoder with a discrete-time flow map that advances the latent variable by
one step,
$\bm{L}_{n+1}=\Phi(\Delta t_n,\bm{L}_n;\theta)$, trained by one-step teacher
forcing on the native non-uniform step sequence, with mass conservation imposed
as a soft penalty in the loss. For a controlled comparison we retrain a flow map
of this form on the identical data: the same frozen autoencoder latent variables, the same
train/validation/test split, and a matched network capacity, with the clock
supplied as simple normalized $\log_{10}t$. One-step training of the direct map
is recursively unstable: without a small train-time input-noise regularization
($\sigma=3\times10^{-3}$) the $999$-step rollout diverges on nearly all
validation trajectories, and even with it, training must be stopped by
rollout-validated early stopping while the one-step loss is still
decreasing, so the baseline reported here is the strongest we could construct.

\begin{table}[t]
\centering
\caption{Head-to-head with the retrained flow-map baseline on $300$ test
trajectories. Metrics are computed on each row's evaluation grid and are
therefore grid-weighted (the arc-length grid concentrates points in the
transition, where the error is largest), so comparisons should be read within a
grid; the direct comparison is the two native-grid rows. Diverged rollouts
($|\bm{L}'|$ exceeding $10^{3}$ or non-finite) are excluded from the error
statistics and counted separately. Timings are batched single-GPU inference
under a common protocol (batches of $300$ trajectories, five repetitions
after a warm-up call, single precision, one A100 GPU; the latent rollout only).}
\label{tab:fmnet}
\small
\begin{tabular}{@{}llccccc@{}}
\toprule
model & grid & steps & median rel.\ $L_2$ & median MAE ($n_z$) & diverged & time/traj.\ \\
\midrule
flow map (FMNet) & native   & $999$ & $1.3\times10^{-1}$ & $1.4\times10^{-2}$ & $5/300$ & $600\,\mu$s \\
neural ODE (ours) & native  & $999$ & $1.3\times10^{-2}$ & $1.6\times10^{-3}$ & $0/300$ & $3770\,\mu$s \\
neural ODE (ours) & adaptive & $99$ & $2.4\times10^{-2}$ & $3.8\times10^{-3}$ & $0/300$ & $374\,\mu$s \\
\bottomrule
\end{tabular}
\end{table}

\begin{figure}[t]
\centering
\includegraphics[width=.78\linewidth]{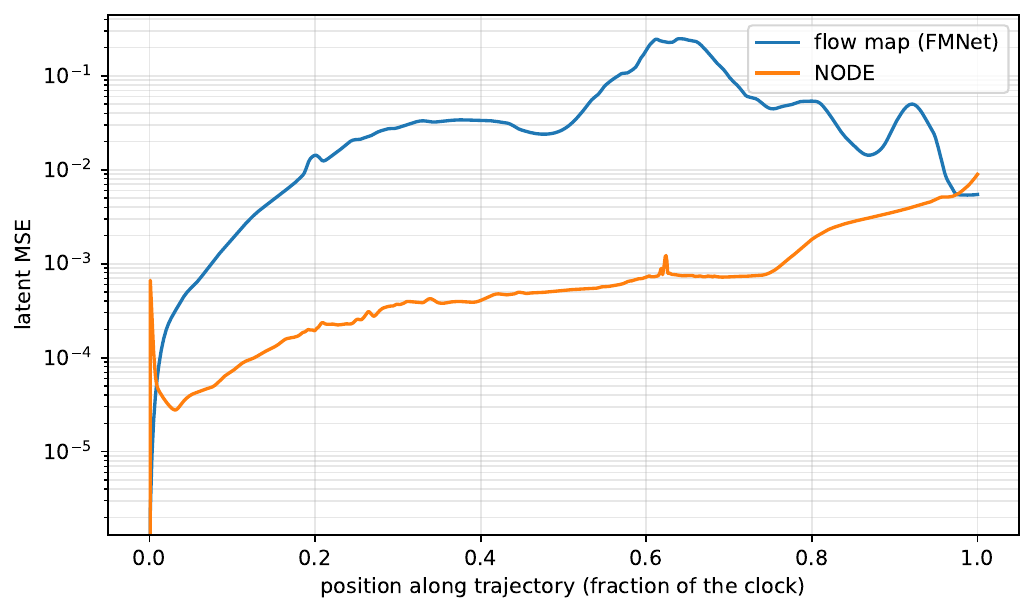}
\caption{Per-step latent error along the full test rollout for the retrained
flow map and the neural ODE, both integrated on the native $1000$-point grid and
plotted against position along the clock. The flow map's error grows by more
than two orders of magnitude over the recursion, the error accumulation
characteristic of one-step-trained discrete maps, while the neural ODE's error
remains locked to the dynamically hard regions of the trajectory. Five of the
$300$ flow-map rollouts diverge and are excluded from its curve; none of the
neural ODE rollouts diverge.} 
\label{fig:fmnet-drift}
\end{figure}

\Cref{tab:fmnet} and \Cref{fig:fmnet-drift} summarize the outcome. On the
identical native grid the continuous-time model is an order of magnitude more
accurate (median relative $L_2$ of $1.3\times10^{-2}$ versus
$1.3\times10^{-1}$; median MAE $1.6\times10^{-3}$ versus $1.4\times10^{-2}$),
its worst-case trajectory is seven times better ($5.9\times10^{-2}$ versus
$4.3\times10^{-1}$), and it exhibits none of the flow map's outright divergences.
The per-step error profile makes the mechanism visible: the flow map's error
compounds along the recursion, whereas the neural ODE's error is a function of
position in the trajectory alone. Per network evaluation the two models cost
about the same (the flow map performs $999$ evaluations against the RK4
integrator's $3996$ on this grid), but the continuous-time model does not need
the fine grid: deployed on its $100$-point arc-length grid it evaluates a full
trajectory with $396$ network calls, faster than the flow map's native-grid
rollout, while retaining its accuracy advantage. These results substantiate,
on matched data and capacity, the limitations that motivated the present work:
the one-step-trained discrete map accumulates error over long rollouts and can
diverge, while the continuous-time formulation does neither on the tested trajectories.

\paragraph{Limitations.}
The study is scoped to a single species (lithium) on a fixed grid of
time-independent background conditions; time-varying $\nA$ and $\Te$, as
occur along a transport-code time step, are not addressed. This is a
structural restriction of the present formulation, not only a missing test:
the clock $\tau$ depends on $\nA$, so a time-varying density changes the clock
itself, and a time-varying $\Te$ changes the conditioning the field was trained
to hold fixed. Coupling to a transport code therefore requires either a
state-defined clock or integration of the learned field in physical time, and
data with time-varying conditions to validate it. The surrogate is
validated within the temporal horizon of the training data and at output
resolutions at or above the training grid; coarser grids under-resolve the
transition, and beyond the trained clock range the surrogate fails
(\Cref{sec:res-field}) and must not be used.
{The anchor network is only coarsely accurate, although this does not
affect rollout accuracy within the horizon (\Cref{sec:setup}); the damping
term is an inductive bias that gave no measurable accuracy gain in a single-run
ablation, and the terminal state is not a stable fixed point of the learned
field (\Cref{sec:res-field}).} Species-number conservation of the decoded populations is
the only quantity enforced exactly; energy balance, reaction rates, and
radiation are learned, not enforced, and the learned latent field is a
predictive model over the sampled trajectories rather than an identified
kinetic law (\Cref{sec:node}). Three further restrictions follow from how the surrogate is
built and tested. First, the main results use a split by trajectory, at
trained conditions; the held-out-condition test of \Cref{sec:res-deploy} covers
interpolation between trained grid points for one random draw, not
extrapolation beyond the trained range of $(\nA,\Te)$. Second, the main
results are integrated on arc-length grids built from the reference
trajectory; the deployable two-pass grid of \Cref{sec:res-deploy} nearly
matches them at twice the cost, whereas grids fixed in advance fail on a few
trajectories, and the $100$-point RK4 rollout is not a converged solution of
the learned field on every trajectory. Third, the clock $\tau$ is elapsed time from
the common initial time $t=10^{-16}\,$s, so the surrogate advances a
trajectory from its initial state but cannot be restarted from an arbitrary
intermediate state, which has no $\tau$ assigned to it. A state on a
computed trajectory restarts at its own clock value with unchanged accuracy,
whereas resetting the clock fails (\Cref{sec:res-field}). Assigning $\tau$ from
the state itself, for example through its charge-state progress, which the
folded clock may make well defined, is a natural remedy.
Several evaluations that would sharpen these statements have not been
run and are left for a later study: the constrained field
$\alpha(\tau)g_\theta(\Lat';c)$ of \Cref{sec:clock}; repetitions of the
ablations of \Cref{sec:res-deploy} over random seeds; a field whose terminal
state is attracting, for example through a constraint on its Jacobian there;
states changed by transport between steps; and validation outside the trained
range of conditions.

\section{Conclusions}
\label{sec:conclusions}
We have presented a continuous-time surrogate for stiff collisional--radiative
kinetics, built on the grey-box latent space of \citet{Xie2024} and driven by a
latent neural ODE with the linear damping of \cite{Linot2023}. The central enabler is a physics-based time
reparametrization: integrating in a density-folded logarithmic clock maps the
sixteen decades of physical time onto a unit interval on which the latent
trajectory is smooth apart from one localized transition, which
arc-length-adaptive sampling resolves, and aligns the ionization
transition across the
parameter grid. The clock does not remove stiffness in the spectral sense;
it is the combination of latent reduction, reparametrization, and adaptive
sampling that lets an explicit Runge--Kutta integrator replace the implicit
stiff solver on the tested trajectories, with a single small vector field ($5.9\times10^{4}$ parameters)
capturing the dynamics of the full ensemble. The integrated field includes a
linear damping toward an estimate of
the CR steady state, regressed on $(\nA,\Te)$ from the simulated terminal states,
as an inductive bias, and the rollouts reach the observed
terminal states at the end of the trained clock interval. The terminal
state is not attracting for the learned field, and a model trained without the
damping is as accurate at the median, so the damping does not produce that
behaviour; the surrogate must not be integrated past the clock range of the
data.

On $1480$ held-out trajectories the surrogate reproduces the charge-state
evolution with a median absolute error of $4.2\times10^{-3}$ on fractions in
$[0,1]$, within a factor of $2.6$ of the autoencoder's own reconstruction baseline;
conserves mass by construction, to within $4\times10^{-7}$; has a
charge-state error that is unchanged under refinement of its output grid at
or above the training resolution, with a per-step latent error that
depends on position along the trajectory rather than on the number of steps
through the ionization transition, but grows with refinement on the late
plateau; and evaluates in $76\,\mu$s per full trajectory on a single GPU in batched inference.
A model retrained with whole $(\nA,\Te)$ grid points held out is as
accurate at those unseen conditions as at trained ones (median MAE
$3.8\times10^{-3}$ and $3.7\times10^{-3}$), and a two-pass grid built from the
model's own prediction replaces the reference grid at nearly the same accuracy
and $153\,\mu$s per trajectory.

The study is scoped to a single species (lithium) on a fixed condition grid, and
its quantitative baseline is a retrained flow-map surrogate rather than an
operator-learning alternative, a comparison that awaits a common CR benchmark.
Natural next steps are a validation beyond the trained range of
$(\nA,\Te)$, a state-defined
clock that allows the surrogate to be restarted from an arbitrary state, the
coupling of the surrogate into an integrated transport
code, where its continuous-time formulation and hard conservation are designed to
matter, and the extension to further species and time-dependent background conditions, for which high-fidelity CR modeling and data are now becoming available,
including hybrid fine-structure and superconfiguration schemes for helium,
lithium, and beryllium \citep{sharma2026hybrid} and CR data for
disruption-mitigation species \citep{sharma2026collisional}.

\section*{Data availability statement}
The data that support the findings of this study are available from the corresponding author upon reasonable request.

\section*{Acknowledgment}
This research used resources of the National Energy Research Scientific Computing Center (NERSC), a Department of Energy Office of Science User Facility, using NERSC award DDR-ERCAP0035032. QT was partially supported by the U.S. Department of Energy’s National Nuclear Security Administration (NNSA) under the Predictive Science Academic Alliance Program (PSAAP-IV), Award DE-NA0004265.

\bibliography{refs}

\end{document}